\documentclass[twocolumn,showpacs,nofootinbib,aps,superscriptaddress,eqsecnum,prd,notitlepage,showkeys]{revtex4-1}

\usepackage{epsf}  
\usepackage{eucal} 
\usepackage{amssymb}
\usepackage{amsmath} 
\usepackage{graphicx} 
\usepackage{bm}
\usepackage{dcolumn} 
\usepackage{epsfig} 
\usepackage{multirow}
\usepackage{hyperref}
\usepackage{subfigure}
\usepackage{amscd}
\usepackage{color} 

\def\jcap{JCAP}%

\def\Phib{\bar{\Phi}}

\newcommand{\mrm}[1]{\mathrm{#1}}
\newcommand{\mc}[1]{\mathcal{#1}}

\newcommand{\blr}[1]{\left(#1\right)}
\newcommand{\be}{\begin{equation}}
\newcommand{\ee}{\end{equation}}
\newcommand{\eea}{\end{eqnarray}}
\newcommand{\bse}{\begin{subequations}}
\newcommand{\ese}{\end{subequations}}
\definecolor{orange}{rgb}{1,0.5,0}

\newcommand{\dirac}{\partial\llap{$\diagup$\kern-2pt}}

\def\be{\begin{equation}} 
\def\ee{\end{equation}}
\def\bq{\begin{eqnarray}} 
\def\eq{\end{eqnarray}}

\begin{document}
           

\title{Phase diagram and surface tension in the three-flavor Polyakov-quark-meson model}

\author{B.~W.~Mintz}\email{bruno.mintz.uerj@gmail.com}
\affiliation{Departamento de F\'{\i}sica Te\'orica, Universidade do Estado 
do Rio de Janeiro, 20550-013, Rio de Janeiro, RJ, Brazil}

\author{R.~Stiele}\email{r.stiele@thphys.uni-heidelberg.de}
\affiliation{Institut f\"{u}r Theoretische Physik, Ruprecht-Karls-Universit\"at Heidelberg,
Philosophenweg 16, D-69120, Heidelberg, Germany}

\author{Rudnei O.~Ramos}\email{rudnei@uerj.br}
\affiliation{Departamento de F\'{\i}sica Te\'orica, Universidade do Estado 
do Rio de Janeiro, 20550-013, Rio de Janeiro, RJ, Brazil}

\author{J.~Schaffner-Bielich}\email{schaffner-bielich@uni-heidelberg.de}
\affiliation{Institut f\"{u}r Theoretische Physik, Ruprecht-Karls-Universit\"at Heidelberg,
Philosophenweg 16, D-69120, Heidelberg, Germany}
\affiliation{{Institut f\"{u}r Theoretische Physik, Goethe-Universit\"at Frankfurt,
Max-von-Laue-Stra\ss e 1, D-60438, Frankfurt am Main, Germany}}
\affiliation{ExtreMe Matter Institute EMMI, GSI Helmholtzzentrum f\"ur
Schwerionenforschung GmbH,
 Planckstra\ss e 1, D-64291 Darmstadt, Germany}

\begin{abstract}

We obtain the in-medium effective potential of the three-flavor
Polyakov-quark-meson model as a real function of real variables
in the Polyakov loop variable, to allow for the study of all
possible minima of the model. At finite quark chemical potential, the
real and imaginary parts of the effective potential, in terms of the
Polyakov loop variables, are made apparent, showing explicitly the
fermion sign problem of the theory. The phase diagram and other
equilibrium observables, obtained from the real part of the effective
potential, are calculated in the mean-field approximation. The
obtained results are compared to those found with the so-called
saddle-point approach. Our procedure also allows the calculation of
the surface tension between the chirally broken and confined phase,
and the chirally restored and deconfined phase. The values of surface
tension we find for low temperatures are very close to the ones
recently found for two-flavor chiral models. Some consequences of our
results for the early Universe, for heavy-ion collisions, and for
proto-neutron stars are discussed.

\end{abstract}



\maketitle

\section{\label{sec1}Introduction}

The determination of the properties of strongly interacting matter at
high temperatures and baryon densities is one of the main goals of
today's high-energy physics. However, not many tools are available for
such a difficult task. Due to the fermion sign problem of quantum
chromodynamics (QCD) at nonvanishing (real) quark chemical potential,
Monte-Carlo calculations on the lattice are not feasible in this
regime due to the lack of an importance sampling procedure that is
free of ambiguities~\cite{deForcrand_06,DeTar_09}. (Nevertheless, 
the authors of \cite{Endrodi:2011gv} have been able to investigate the 
curvature of the $T-\mu$ phase diagram at low values of chemical potential 
on the lattice, with physical quark masses and in the continuum extrapolation, 
indicating that the crossover temperature decreases with the chemical 
potential for $\mu/T_c\ll1$.) An alternative approach is that of chiral 
effective models for QCD, which have been successfully utilized for many 
decades~\cite{Nambu_61_A,GellMann_60}. In this work, we shall adopt one 
such model, the three-flavor {\it Polyakov-quark-meson} (PQM)
model~\cite{Schaefer_07,Schaefer_10,Herbst_11, Schaefer_12}. It has
become quite popular in the last years due to its close relationship
with the linear sigma model and its agreement with results from
lattice calculations of thermodynamical quantities at zero baryon
chemical potential \cite{Haas:2013qwp, Stiele:2013}, most importantly 
the crossover nature of the quark-hadron transition at vanishing chemical 
potential \cite{Aoki:2006we}. 

As discussed for example
in~\cite{Fukushima_07,Dumitru_05,Roessner_08}, not only QCD, but also
effective models that possess gauge degrees of freedom do present
the sign problem at finite chemical potential $\mu$, even at the
mean-field level. In such models, this is manifest in the appearance
of an imaginary part of the in-medium effective potential in
equilibrium at $\mu \neq 0$. Since the resulting effective potential
at finite chemical potential is a complex function of complex
variables, special care must be taken with respect to the meaning of a
minimization procedure that leads to the state of equilibrium of the
system at any given nonzero temperature $T$ and chemical potential
$\mu$. Following an approach similar to the one
in~\cite{Roessner_08} (in the context of the PNJL model), we propose a
change of variables, followed by a simple approximation in the PQM
model that renders the in-medium effective potential a real function
of real variables. As a consequence, the effective potential is, in
this approach, a real function that possesses minima, as demanded by
general field-theoretical arguments applied to systems in
equilibrium. 

The implementation of such an approximation scheme (that simply
accounts for neglecting the imaginary part of the effective potential)
of course leads to differences in the predictions made, for example,
with the so-called {\it saddle-point}
approach (see, e.g., \cite{Fukushima_04}). In this approach, the state
of thermodynamical equilibrium is found by first restricting the
Polyakov loop variables to real variables. Second, all the extrema of
the effective potential ({\it i.e.}, points in which the derivatives
of the effective potential with respect to all order-parameters
vanish) are determined. At a third step, the state of equilibrium is
chosen among the extrema as the one with the lowest value of the
effective potential. It often happens that the chosen point is not a
minimum, but a saddle-point of the effective potential. In fact, in
this approach, the effective potential often has no minima, but only
saddle-points. 

In this work, we show how the differences between the two approaches
are manifest in the $T-\mu$ phase diagram of the model. 
We also briefly discuss differences
and similarities between several parametrizations of the Polyakov loop
potential found in the literature (see, e.g.,
\cite{Scavenius_02,Ratti_06,Roessner_07}). The saddle-point approach is
particularly troublesome for the computation of the surface
tension in the region of the phase diagram of the model where a
first-order phase transition exists. The computation
of the surface tension, and thus of nucleation rates, requires a
precise definition and location of the minima of the model. This
adds another important aspect to our suggested approach, which is
that it consistently allows the evaluation of the surface tension
between two phases of the model, which are separated by a first-order
phase transition. As discussed for example
in~\cite{Langer_73,Csernai:1992tj,Gleiser:1993hf,Scavenius_01,Bessa_09,
Bombaci:2008wg,Bombaci:2009jt,Mintz_10,Boeckel_10, Boeckel_12},
the surface tension is a crucial parameter for the dynamics of a
first-order phase transition in many scenarios of high-energy
physics. Roughly speaking, if the surface tension between the stable
and the metastable phases in a given phase transition is too high,
the phase conversion may be exceedingly slow. In practical terms, a
high surface tension can dynamically suppress a first-order phase
transition that would be otherwise allowed if only the bulk equilibrium
thermodynamics were considered. For this reason, the evaluation of
the surface tension from effective models of QCD is a relevant
question about strongly interacting systems that can undergo a
first-order phase transition, such as the ones produced in
relativistic heavy-ion collisions at RHIC, NICA and FAIR, and also in
the interior of compact stars. This evaluation, however, is not
available from first-principles QCD but can be performed in a
relatively straightforward manner using an effective model for QCD
(with or without Polyakov loop degrees of freedom), such as the
two-flavor quark-meson model~\cite{Palhares_10}, the Nambu-Jona Lasinio (NJL)
model~\cite{Pinto_12}, or the PQM model, which we will be studying in
this work. Another goal in this work is to produce a result that at
least overestimates the surface tension in the three-flavor PQM model
along the line of first-order phase transition in the $T-\mu$
plane. This is an interesting result by itself, but it is also an
interesting example of a calculation of the surface tension in a
theory with multiple order-parameters. Notice that the calculation of
the surface tension explicitly demands not only the localization of
the minima of the effective potential, but also the precise form of
the potential between them. We argue that having a real effective
potential as a function of real order-parameters is crucial for the
calculation of the surface tension, as well as for any other quantity
in full thermodynamical equilibrium.

The structure of the paper is as follows. After a brief summary of the
PQM model, in Sec.~\ref{sec2}, we will discuss the necessary
conditions on the effective potential for the definiteness of
thermodynamical equilibrium states and how these apply to the
model. We also show how the sign problem explicitly appears in the PQM
model at the mean-field level and propose an approximation in order
to circumvent it. In Sec.~\ref{sec3}, for completeness, we start the
section by reviewing the general computation of surface tensions in
a first-order phase transition. After that, we discuss how to obtain
an overestimate of the surface tension of a phase interface in the PQM
model. In
Sec.~\ref{sec4} we show our results for the phase diagram for the 
PQM model and the
surface tension. The implications of these results are also discussed.
Finally, in Sec.~\ref{sec5}, we give our
conclusions and final discussions of the obtained results.

\section{\label{sec2}The Polyakov-quark-meson Model at Finite $T$ and $\mu$}

In order to incorporate aspects from the physics of chiral symmetry
breaking and restoration, as well as from the
confinement-deconfinement phase transitions, we adopt an effective
model that captures these main features of quantum chromodynamics.
In the last years, two such models have gained much
popularity, the Polyakov-loop extended Nambu-Jona Lasinio (PNJL) model
\cite{Mocsy_04,Fukushima_04,Megias_06_A,Ratti_06} and the Polyakov-loop-extended
quark-meson (PQM) model
\cite{Schaefer_07,Schaefer_10,Herbst_11,Schaefer_12}. These models may
be considered as extensions of the well-known Nambu-Jona Lasinio
model \cite{Klevansky_92,Buballa_05,Costa_08} and the linear sigma
model
\cite{Metzger_93,Lenaghan_00,Scavenius_00,Scavenius_01,Bowman_08,Schaefer_08},
which provide an effective realization of the chiral symmetry breaking pattern. 
The physics of confinement is expected to be
taken into account by coupling the quark and the meson degrees of
freedom with the expectation value of the Polyakov loop.

In this work, our choice will be a PQM model with $2+1$ quark flavors,
as in Refs.~\cite{Schaefer_10,Schaefer_12}.

\subsection{The Polyakov loop}

Before discussing the model, we briefly introduce the order
parameter for (de)confinement with which the sign problem enters
in Polyakov-loop extended models.

The Polyakov loop operator is a Wilson loop in temporal direction,
\be
 	\mc{P}=P \exp \blr{ i \int_0^\beta dx_0 \,A_0\blr{x_0}}\;,  
\ee
where $P$ denotes path ordering, $\beta=1/T$ is the inverse
of the temperature and $A_0$ is the temporal component of the gauge field
$A_\mu$ \cite{Polyakov_78}.
With a gauge that ensures the time independence of $A_0$,
we can perform the integration trivially and the path ordering
becomes irrelevant \cite{Marhauser_08,Megias_06_B}, so that $\mc{P}=\exp \blr{ i \beta A_0}$.
In this form, it is trivial to see that the Polyakov loop variable,
\be
	\Phi = \frac{1}{N_\mrm{c}} \mrm{tr}\, \mc{P}
\ee
is a complex scalar field $	\Phi = \alpha + i \beta$.
Furthermore, we can rotate the gauge field in the cartan subalgebra
$A_0^c = A_0^{(3)} \lambda_3 + A_0^{(8)} \lambda_8$ \cite{Braun:2007bx}.
Within this diagonal representation, we see that the adjoint Polyakov loop
variable becomes simply $\Phib = \alpha - i \beta$.

Under gauge transformations concerning center symmetry,
the Polyakov loop operator and its variable are multiplied with
a center element $\mc{Z}$, $\Phi \to \mc{Z}\Phi$.
In pure $SU(3)$ gauge theory, that can be considered as QCD with
infinitely heavy quark masses, the confining phase is center-symmetric
and, therefore, $\langle\Phi\rangle = 0$, while deconfinement is characterized
by a finite value of the Polyakov loop expectation value, since center
symmetry gets broken spontaneously \cite{Polyakov_78}. Real QCD with
physical quark masses adds an explicit symmetry breaking term.

\subsection{Formulation of the Model}

Let us briefly recall a few aspects of the PQM model. The interested
reader may refer to e.g.
Refs.~\cite{Lenaghan_00,Schaefer_08,Schaefer_10,Schaefer_12} for more
details.

The starting point for our analysis of the equilibrium states of the
system is the in-medium effective potential of the theory\footnote{As
long as we are interested in states of thermal equilibrium of the
system, which are described by homogeneous field configurations, the
in-medium effective action is reduced to an in-medium effective
potential. In the next section, where we study non-homogeneous
states, we will have to consider the effective action instead.} as
a function of the relevant order-parameters and thermodynamical
control parameters. This effective potential is derived within a
mean-field approximation to a theory with constituent quarks
minimally coupled to gauge fields and coupled to mesons via a
Yukawa-type term. In the leading in-medium contribution from the
fermions, the coupling to the color fields effectively becomes a
coupling to the Polyakov loop field. The coupling with the mesons is
translated into the masses of the quarks via spontaneous and explicit
chiral symmetry breaking terms in the self-interaction Lagrangian
density for the mesons. The self-interaction in the gauge sector is
modeled by a potential energy for the Polyakov loop variable. 

In the chiral sector of the theory, the natural choice of order
parameters for the $N_f=2+1$ pattern of chiral symmetry breaking are
the non-strange ($\sigma_x$) and strange ($\sigma_y$) chiral
condensates~\cite{Schaefer_08}. In the Polyakov-loop sector, the
natural variables are the Polyakov loop itself, $\Phi$, and its
conjugate, $\Phib$. With these choices, the effective potential is
given by the sum of three terms,
\be\label{eq:grand_canon_pot} \Omega = U(\sigma_x, \sigma_y) + {\cal
 U}(\Phi,\bar{\Phi}) +
\Omega_{q\bar{q}}(\sigma_x,\sigma_y,\Phi,\bar{\Phi}) \ee where the
dependence on the thermodynamical control parameters (in this case the
temperature $T$ and the quark chemical potential $\mu$) is implicit
in the last two terms of Eq.~(\ref{eq:grand_canon_pot}). 
Below, we describe in detail each of the contributions appearing in the
thermodynamical potential Eq.~(\ref{eq:grand_canon_pot}).

The first term of the thermodynamical potential
(\ref{eq:grand_canon_pot}) is the tree-level contribution from the
mesonic degrees of freedom and reads
\begin{eqnarray}\label{eq:meson_tree_level_pot}
	U\blr{\sigma_x,\sigma_y} &=& \frac{m^2}{2}\blr{\sigma_x^2 +
          \sigma_y^{2}} + \frac{\lambda_1}{2} \sigma_x^2 \sigma_y^2 +
        \nonumber\\ && + \frac{1}{8}\blr{2 \lambda_1 +
          \lambda_2}\sigma_x^4+\frac{1}{8}\blr{2 \lambda_1 +
          2\lambda_2} \sigma_y^4 -\nonumber\\ && - \frac{c}{2
          \sqrt{2}} \sigma_x^2 \sigma_y -h_x \sigma_x - h_y \sigma_y
        \;.
\end{eqnarray}

Since we are studying isospin symmetric matter, we do not distinguish
the up and down quark sectors. The mesonic sector has six parameters,
that are the mass and couplings $m^2$, $\lambda_1$, $\lambda_2$ and
$c$ and the explicit chiral symmetry breaking terms, $h_x$ and
$h_y$. They are adjusted to the pion and kaon decay constants $f_\pi$
and $f_K$ and meson masses of the scalar and pseudoscalar octet,
$m_\pi$, $m_K$, $m_\eta^2+m_{\eta'}^2$ and $m_\sigma$. The mass of the
sigma meson is still a poorly known number, but the most recent
compilation of the Particle Data Group~\cite{PDG_12} considers that
$m_\sigma$ can vary between $400\,\mrm{MeV}$ and
$550\,\mrm{MeV}$. Once given this set of masses and decay constants,
the model parameters are defined. For their explicit expressions,
please refer to Refs.~\cite{Lenaghan_00,Schaefer_08}. The values of
the constants we use to calculate these parameters are listed in
Table~\ref{tab:chiral_pot_constants}. We will be using frequently 
the value $m_\sigma=500~$MeV for the sigma mass. 

\begin{table}
	\caption{Values of constants to which the parameters of the
          mesonic potential are adjusted, according to
          Ref.~\cite{PDG_12} and value of the {\it constituent} quark
          mass of the light (up and down) quarks that we use to fix
          the quark meson Yukawa coupling in
          Eq.~(\ref{eq:quarkmasses_coupling}).}
	\begin{ruledtabular}
		\begin{tabular}{l||ccccccc|c}
			Constant & $f_\pi$ & $f_K$ & $m_\pi$ & $m_K$ &
                        $m_\eta$ & $m_{\eta'}$ & $m_\sigma$ & $m_l$
                        \\ \hline  Value [MeV] & 92 & 110 & 138 & 495
                        & 548 & 958 & $400-550$ & 300 \\
		\end{tabular}
	\end{ruledtabular}
	\label{tab:chiral_pot_constants}
\end{table}

The second term in Eq.~(\ref{eq:grand_canon_pot}) is responsible for
including the physics of color confinement through the introduction of
a potential energy for the expectation value of the Polyakov loop. The
explicit functional form of the potential energy density for the
Polyakov loop is still not known directly from first-principle
calculations. Instead, a common approach is to choose a functional
form for the potential that reproduces crucial features of pure gauge
theory and then adjust a set of free parameters to the results for the
Polyakov loop and the thermodynamical observables of Monte-Carlo
lattice calculations. A possible parametrization of the Polyakov loop
potential is the {\it polynomial parametrization}~\cite{Ratti_06,
Schaefer_07},
\begin{eqnarray}\label{eq:Polyakov_potential_polynomial}
	\frac{\mc{U}_{\text{poly}}\blr{\Phi,\Phib}}{T^{4}} &=&
        -\frac{b_2\blr{T}}{2} \Phib\Phi -
        \frac{b_3}{6}\blr{\Phi^3+\Phib^3} + \nonumber\\ && +
        \frac{b_4}{4} \blr{\Phib\Phi}^2
\end{eqnarray}
with the temperature-dependent coefficient $b_2$ defined as
\be
	b_2\blr{T} = a_0 + a_1 \blr{\frac{T_0}{T}} + a_2 \blr{\frac{T_0}{T}}^2 + a_3 \blr{\frac{T_0}{T}}^3\ . 
\ee 
An equivalent polynomial parametrization had been previously proposed
in~\cite{Scavenius_02}, with different definitions for the
coefficients in Eq.~(\ref{eq:Polyakov_potential_polynomial}). A simple
calculation allows the translation from one set of coefficients to
the other. As far as the polynomial parametrization is concerned, we
shall stick to the form (\ref{eq:Polyakov_potential_polynomial}).

Another possible parametrization for the effective potential of the
Polyakov loop is provided by \cite{Roessner_07}
\begin{multline}\label{eq:Polyakov_potential_logarithmic}
	\frac{\mc{U}_{\text{log}}\blr{\Phi,\Phib}}{T^{4}} =
        -\frac{1}{2}A \blr{T} \Phib\Phi + B\blr{T} \times  \\ \times
        \ln \left[ 1 - 6 \blr{ \Phib\Phi} + 4 \blr{\Phi^{3}+\Phib^{3}}
          - 3\blr{ \Phib\Phi}^{2}\right],
\end{multline}
where both coefficients are temperature dependent, 
\bse
\begin{eqnarray}
	A\blr{T} &=&  A_0 + A_1 \blr{\frac{T_0}{T}} + A_2
        \blr{\frac{T_0}{T}}^2,\\ B\blr{T} &=& B_3
        \blr{\frac{T_0}{T}}^3.
\end{eqnarray}
\ese The form~(\ref{eq:Polyakov_potential_logarithmic}) of the
Polyakov loop potential is called the {\it logarithmic
 parametrization}. In Refs~\cite{Scavenius_02,Ratti_06, Roessner_07} the
parameters of both the polynomial and of the logarithmic
parametrizations (see next) were adjusted to the lattice simulation
of Ref.~\cite{Boyd_96}. They are listed in
Table~\ref{tab:Ploop_pot_params}.

\begin{table}
	\caption{Parameters of the gauge potential parametrizations
          for fits to the lattice simulation \cite{Boyd_96}.}
	\begin{ruledtabular}
		\begin{tabular}{lcccccc}
			poly         & $a_0$ & $a_1$ & $a_2$ & $a_3$ &
                        $b_3$   & $b_4$ \\ \cite{Scavenius_02}  & 1.53
                        & 0.96  & -2.3  & -2.85 & 13.34   & 14.88
                        \\ \cite{Ratti_06}      & 6.75  & -1.95 &
                        2.625 & -7.44 & 0.75    & 7.5 \\
			\hline
			log & $A_0$ & $A_1$ & $A_2$ & $B_3$
                        \\ \cite{Roessner_07} & 3.51 & -2.47 & 15.2 &
                        -1.75  \\
		\end{tabular}
	\end{ruledtabular}
	\label{tab:Ploop_pot_params}
\end{table}
Originally, the parameter $T_0$ was devised to correspond to the
transition temperature in pure Yang-Mills theory
$T_0=270\,\mrm{MeV}$. However, in full dynamical QCD, fermionic
contributions and the matter backreaction modify the pure gauge
potential to an effective glue potential. Reference~\cite{Schaefer_07}
estimated this running coupling of QCD by consistency with hard
thermal loop perturbation theory calculations \cite{Braun_06,
 Braun_07}. They mapped this effect to an $N_f$-dependent
modification of the expansion coefficients of the Polyakov loop
potential that results in 
a $N_f$-dependence of $T_0$.
The actual value of $T_0$ for $2+1$ quark flavors with a {\it current} 
strange quark mass of 95\,MeV \cite{PDG_12} is $T_0=182\,$MeV.

The authors in Ref.~\cite{Schaefer_07} also estimated the dependence
of the glue potential with the quark density and mapped it to a quark
chemical potential dependence of $T_0$. Such a dependence can be
expected in view of a $\mu$-dependent color screening effect due to
quarks. The $\mu$ dependence of $T_0$ suggested in \cite{Schaefer_07}
is implemented as a small correction to the running coupling. Here, we
will simply take a more simplistic approach, but without loss of
generality, and consider $T_0$ as a
constant parameter.

An improved mapping between the pure Yang-Mills potential and the quark-improved 
glue effective potential in full QCD is discussed and applied in 
Refs.~\cite{Haas:2013qwp, Stiele:2013} and will be taken into account 
in a future work.

Finally, the last term of Eq.~(\ref{eq:grand_canon_pot}) represents
the constituent quark sector coupled to the gauge field (represented
by the Polyakov loop variables) and to the mesons, 
\begin{multline}\label{eq:O_qqb}
	\Omega_{q\bar{q}}(\sigma_x,\sigma_y,\Phi,\Phib)=
-2T\sum_{f=u,d,s}\int\frac{d^3p}{(2\pi)^3}\times\\ \times\left\{\ln\left[
          1 + 3\blr{\Phi + \Phib e^{-\blr{E_{q,f}-\mu_f}/T}}\times
          \right.\right. \\ \left.\times e^{-\blr{E_{q,f}-\mu_f}/T} +
          e^{-3\blr{E_{q,f}-\mu_f}/T}\right] + \\ + \ln\left[1 +
          3\blr{\Phib + \Phi e^{-\blr{E_{q,f}+\mu_f}/T}}\times
          \right. \\ \left.\left.\times e^{-\blr{E_{q,f}+\mu_f}/T} +
          e^{-3\blr{E_{q,f}+\mu_f}/T}\right]\right\}\;,\qquad
\end{multline}
where $\mu_f$ is the quark chemical potential for each of the quark
flavors.

The constituent light and strange quark masses are, respectively,
\begin{equation}\label{eq:quarkmasses_coupling}
	m_l = \frac{g}{2} \sigma_x \qquad \text{and} \qquad m_s =
        \frac{g}{\sqrt{2}} \sigma_y\;.
\end{equation}
To fix the Yukawa coupling, we choose the constituent quark mass of
the light (up and down) quarks to be $m_l = 300\,\mrm{MeV}$ in the
vacuum, where $\langle\sigma_x\rangle=f_\pi$. This results in $m_s
\simeq 417\,$MeV for the constituent strange quarks.

Notice that the dependence of $\Omega_{q\bar q}$ with the chiral
condensates $\sigma_x$ and $\sigma_y$ in Eq.~(\ref{eq:O_qqb}) is
implicit in the quasi-particle dispersion relation for the constituent
quarks
\begin{equation}
	E_{q,f} = \sqrt{k^2+m_f^2}\;.
\end{equation}

One last comment should be made about the vacuum (or {\it sea}) terms
of the in-medium effective potential. Many comparative studies within
the NJL and linear sigma models and comparisons with functional 
calculations \cite{Braun_11, Herbst_11, Pawlowski:2010ht} have shown that vacuum terms can
change qualitatively the structure of the $T-\mu$ phase diagram. In
particular, if the sigma meson mass is sufficiently large, the line
of first-order phase transition can eventually disappear due to the
effect of the fermionic vacuum contribution at one-loop
\cite{Bowman_08,Skokov_10,Andersen_11}. In the present work, although
we neglect the vacuum terms from both fermionic and bosonic fields, we
adopt mainly a low value of the sigma mass, $m_\sigma=500\,\mrm{MeV}$. Such a
low value has been shown to allow for a first-order transition at
low temperatures even with the inclusion of fermionic vacuum
contributions only (see for instance Ref.~\cite{Schaefer_12} and
references therein). Therefore, we believe that the no-sea
approximation does not spoil the general conclusions of this
work. Still, we agree with the authors of \cite{Andersen_11} that {\it
 all} the contributions from all fields of a model should be
consistently taken into account, that is the case in functional calculations 
\cite{Braun_11, Pawlowski:2010ht}. In this direction, a full study of
the renormalized one-loop contributions from quarks and mesons in the
PQM model will be taken into account in a future work
\cite{MintzRamos_13}.

\subsection{Thermodynamical equilibrium}

Given the temperature $T$ and the quark chemical potential $\mu$,
the effective potential (\ref{eq:grand_canon_pot}) is then given as a
function of the four order-parameters of the model, $\sigma_x$,
$\sigma_y$, $\Phi$ and $\Phib$. In thermal equilibrium, the field
configurations that contribute the most to the partition function are
those that {\it minimize} the in-medium effective potential (or the
in-medium effective action, for non-homogeneous systems). All other
extrema of the effective action are exponentially suppressed and give
negligible contributions to the equilibrium thermodynamics of the
system. 

Even though this issue has already been addressed before in the
literature (e.g. in Refs.~\cite{Fukushima_07,Roessner_08}), we believe
that it should be discussed more thoroughly, especially in the
context of the PQM model. Our main motivations for pressing on this
point are, first, theoretical consistency, second, that the results from
the two approaches (the saddle-point approach and the present one) are
different, and third, that only with equilibrium states described by
minima of the effective potential can one calculate quasi-equilibrium
properties of the system, such as the surface tension in a
first-order phase transition, as we already commented above in
Sec.~\ref{sec1}. 

The one-loop effective potential of the PQM model at finite
temperature and quark chemical potential, as defined in
Eq.~(\ref{eq:grand_canon_pot}), is a complex function of complex
variables. 
Therefore, it can have no minima and, as it stands, it
cannot provide a standard description for a thermodynamical system in
equilibrium. Finally, it makes sense to identify the extrema of a
function with maxima, minima and saddle-points only if it is a real
function of real variables.

In order to make this statement clearer, let us explicitly identify
the real and imaginary parts of the effective potential at finite $T$
and $\mu$.\footnote{The following discussion will make it clear that 
at $\mu=0$, $\Phi$ and $\Phib$ are identical and real, so that there is
 no sign problem in that case.} Thus, we must write down the effective potential in terms
of these real variables only. 

Let us start by making a change of variables in the potential by
introducing the real and imaginary parts of the Polyakov loop
variables as
\begin{equation}\label{eq:def_alpha_beta}
 \alpha \equiv\frac{\Phi+\Phib}2\;\;\;\mbox{and}\;\;\;
\beta \equiv \frac{\Phi-\Phib}{2i}.
\end{equation}
The polynomial potential (\ref{eq:Polyakov_potential_polynomial}) and
the logarithmic potential (\ref{eq:Polyakov_potential_logarithmic}),
respectively, can be both rewritten as 
functions of the real
variables $\alpha$ and $\beta$ as
\begin{equation}\label{eq:Ploop_pot_poly_alphabeta}
 \frac{\bar{\cal U}_{\mrm{poly}}(\alpha,\beta)}{T^4} =
 -\frac{b_2}2(\alpha^2+\beta^2) - \frac{b_3}3(\alpha^3 -
 3\alpha\beta^2) + \frac{b_4}4(\alpha^2 + \beta^2)^2,
\end{equation}
and
\begin{multline}\label{eq:Ploop_pot_log_alphabeta}
  \frac{\bar{\cal U}_{\mrm{log}}(\alpha,\beta)}{T^4} =
  -\frac{A(T)}2(\alpha^2+\beta^2) \\  + B(T)\log\!\left[
    1\! -\! 6(\alpha^2 \!+\!\beta^2)\! + \!8(\alpha^3 -
    3\alpha\beta^2) - 3(\alpha^2 + \beta^2)^2\right].
\end{multline}

We now turn to the contribution from the quarks at finite temperature
and chemical potential. Once $\Phi$ and $\Phib$ are complex-valued
variables, the potential~(\ref{eq:O_qqb}) can take complex
values. Let us rewrite it as a function of real variables to make this
statement explicit.

Dropping the flavor indexes to keep the notation simpler, we define
\begin{equation}
 z_+ \equiv 1 + 3(\Phi + \Phib e^{-(E-\mu)/T} )e^{-(E-\mu)/T} +
 e^{-3(E-\mu)/T} ,
\end{equation}
and
\begin{equation}
 z_- \equiv 1 + 3(\Phib + \Phi e^{-(E+\mu)/T} )e^{-(E+\mu)/T} +
 e^{-3(E+\mu)/T},
\end{equation}
such that (\ref{eq:O_qqb}) can now be written as
\begin{equation}\label{eq:O_qqb_zz}
 \Omega_{q\bar q} = -2T\sum_f\int\frac{d^3p}{(2\pi)^3}\log[z_+z_-].
\end{equation}

After using (\ref{eq:def_alpha_beta}) and performing some
straightforward manipulations, one can see that the argument of the
logarithm in (\ref{eq:O_qqb_zz}) is complex, that is,
\begin{equation}
 z_+z_- = R+iI,
\end{equation}
where
\begin{eqnarray}\label{eq:real_part_arglog_qqbar}
 R \equiv && 1 + e^{-3(E-\mu)/T} + e^{-3(E+\mu)/T} + e^{-6E/T} +
 \nonumber\\ && + 6\alpha e^{-E/T}\left[\cosh\left(\frac\mu T\right) +
   e^{-E/T}\cosh\left(\frac{2\mu}T\right)\right] + \nonumber\\ && +
 6\alpha e^{-4E/T}\left[\cosh\left(\frac{2\mu}T\right) +
   e^{-E/T}\cosh\left(\frac\mu T\right)\right] + \nonumber\\ && +
 9(\alpha^2+\beta^2)(1+e^{-2E/T})e^{-2E/T} + \nonumber\\ && +
 18(\alpha^2-\beta^2)e^{-3E/T}\cosh\left(\frac\mu T\right)\;,
\end{eqnarray}
and
\begin{eqnarray}\label{eq:imaginary_part_arglog_qqbar}
 I \equiv && 6\beta e^{-E/T}\left[\sinh\left(\frac\mu T\right) -
   e^{-E/T}\sinh\left(\frac{2\mu}T\right)\right] + \nonumber\\ && +
 6\beta e^{-4E/T}\left[e^{-E/T}\sinh\left(\frac\mu T\right) -
   \sinh\left(\frac{2\mu}T\right)\right] - \nonumber\\ && -
 36\alpha\beta\sinh\left(\frac\mu T\right)e^{-3E/T}.
\end{eqnarray}

The complex argument of the logarithm can be written in polar form,
$R+iI = \rho e^{i\theta}$, with
\begin{equation}
 \rho \equiv\sqrt{R^2 +
   I^2}\;\;\;\mbox{and}\;\;\;\theta \equiv\arctan\left(\frac IR\right)\;,
\end{equation}
so that the potential can be cast in a manifestly complex form,
\begin{equation}
 \Omega_{q\bar q} = \Omega_{q\bar q}^R + i\,\Omega_{q\bar q}^I \;,
\end{equation}
where
\begin{equation}\label{eq:real_part_omega_qqbar}
 \Omega_{q\bar q}^R\equiv-2T\sum_f\int\frac{d^3p}{(2\pi)^3}\log[\rho] \;,
\end{equation}
and
\begin{equation}\label{eq:imaginary_part_omega_qqbar}
 \Omega_{q\bar q}^I\equiv -2T\sum_f\int\frac{d^3p}{(2\pi)^3}\theta.
\end{equation}

The imaginary part (\ref{eq:imaginary_part_omega_qqbar}) is the
manifestation of the fermion sign problem in the context of the PQM
model already at the one-loop level. Note that it is very closely
related to the sign problem in the PNJL model as well
\cite{DeTar_09,Dumitru_05,Fukushima_07}. An important
aspect of Eq.~(\ref{eq:imaginary_part_omega_qqbar}) is that it
vanishes for $\mu=0$, so that the effective potential becomes real
and free of the sign problem. Furthermore,
(\ref{eq:imaginary_part_omega_qqbar}) is odd in $\beta$, while the
real part (\ref{eq:real_part_omega_qqbar}) is even in $\beta$. This
means that we must have $\langle\beta\rangle=0$, {\it i.e.},
$\langle\Phi\rangle=\langle\Phib\rangle$ for $\mu=0$, which is a well-known result.

\subsection{A Comment about the Sign Problem}

In order to make the sign problem explicit in the PQM model, let us
write the grand partition function for the model in the mean-field
approximation,
\begin{equation}\label{eq:grand_partition_function_PQM}
 {\cal Z}=\int[D\vec\Psi]\exp\left[{-\frac {\cal
       V}T\Omega(\vec\Psi)}\right],
\end{equation}
where ${\cal V}$ is the volume of the space and $\vec\Psi=(\sigma_x,
\sigma_y, \alpha, \beta)$ formally represents the (real) order
parameters. 

Notice from Eqs.~(\ref{eq:imaginary_part_arglog_qqbar}) and
(\ref{eq:imaginary_part_omega_qqbar}) that the imaginary part of the
quark contribution is odd in $\beta$ and all the other contributions
are even in $\beta$. Let us then split the effective potential in
a $\beta$-odd part (that coincides with $\Omega_{q\bar q}^I$) and a
$\beta$-even part. The functional integral is to be performed for
every possible (real) value that $\beta$ can assume. We can organize
the sum such that the contributions for a given $\beta_0$ and its
negative $-\beta_0$ are assembled. After a few simple manipulations,
one finds
\begin{eqnarray}\label{eq:partition_function_sign_problem}
 {\cal Z} \!\!&=&\!\! \int_{\beta\in {\rm
     I\!R}}[D\vec\Psi]\exp\left[{-\frac {\cal
       V}T\left(\Omega_{\beta-even} + i\Omega_{q\bar
       q}^I\right)}\right]\nonumber\\ \!\!&=&\!\!
 \int[D\vec\Psi]\exp\left[-\frac {\cal V}T\Omega_{\beta-even}\right]
 \cos\left[-\frac {\cal V}T\Omega_{q\bar q}^I\right].
\end{eqnarray}

The integrand in the expression
(\ref{eq:partition_function_sign_problem}) for the partition function
is not positive defined, as a sound Boltzmann factor should
be. Recall that the integrand of the partition function (the density
matrix elements) corresponds to a sort of probability density, which
must be non-negative. This is not true for the grand-partition
function (\ref{eq:partition_function_sign_problem}) and we conclude
that the PQM model has the sign problem at the mean-field level for a
finite chemical potential.

If we insist on writing the partition function
(\ref{eq:partition_function_sign_problem}) in the same form as
(\ref{eq:grand_partition_function_PQM}), we end up with 
\begin{equation}
 \hat{\Omega} = \Omega_{\beta-even} - \frac{T}{\cal
   V}\log\left[\cos\left(\frac {\cal V}T\Omega_{q\bar
     q}^I\right)\right],
\end{equation}
where ${\cal V}$ is the volume of space. This function is not
physically acceptable as an effective potential. First, it is a
volume-dependent effective potential (therefore, it is not intensive)
and second, it is not defined in the thermodynamical limit ${\cal
 V}\rightarrow\infty$. 

One possibility to circumvent the sign problem, as indicated
in~\cite{Roessner_08} is to treat the imaginary part of the effective
potential perturbatively in an expansion in powers of $T/{\cal V}$. In the
first order of the approximation (which the authors of
\cite{Roessner_08} identify with the mean-field approximation), one
simply ignores the imaginary part (\ref{eq:imaginary_part_omega_qqbar}) 
of the effective potential at finite chemical potential\footnote{In this 
model, ignoring $\Omega_{q\bar q}^I$ is equivalent to taking the modulus of 
the Dirac determinant that is tacitly present in 
(\ref{eq:partition_function_sign_problem})}. 
However, if one neglects $\Omega_{q\bar
 q}^I$, the expectation value of $\beta$ is zero due to the even
parity of $\Omega_{q\bar q}^R$ with respect to $\beta$. As a
consequence, the difference $\langle\Phib-\Phi\rangle=0$, which is in
disagreement with complex Langevin~\cite{Karsch_85} and Monte-Carlo
simulations~\cite{deForcrand_06,DeTar_09}. In spite of this setback,
we understand that our approach is simply an approximation scheme that
has the strong theoretical advantage of dealing with the
well-established minimization procedure for finding the state of
equilibrium.


\subsection{Finding the Minima of the Effective Potential}

From now on, we shall neglect the imaginary part of the effective
potential, as explained before, {\it i.e.}, consider 
\begin{equation}\label{eq:effective_potential_alphabeta}
{\bar \Omega} = U(\sigma_x,\sigma_y) + \bar{\cal U}(\alpha,\beta) +
 \Omega_{q\bar q}^R(\sigma_x,\sigma_y,\alpha,\beta)\;.
\end{equation}

Given the temperature $T$ and the quark chemical potential $\mu$, the
effective potential (\ref{eq:effective_potential_alphabeta}) is then
given as a function of the four order-parameters of the model,
$\sigma_x$, $\sigma_y$, $\alpha$ and $\beta$. In equilibrium, the
expected values for the order parameters are given by minimizing the
effective potential (\ref{eq:effective_potential_alphabeta}) with
respect to each of them. Therefore, the first necessary condition is
to find a point in the order-parameter space such that
\begin{equation}\label{eq:zero_gradient_condition}
 \frac{\partial{\bar \Omega}}{\partial\sigma_x} = 
\frac{\partial{\bar \Omega}}{\partial\sigma_y}   
 = \frac{\partial{\bar \Omega}}{\partial\alpha}
 = \frac{\partial{\bar \Omega}}{\partial\beta} = 0\;,
\end{equation}
when the derivatives are evaluated at the extrema.

The second condition is the positivity of all the eigenvalues of the
$4\times4$ Hessian matrix
\begin{equation}\label{eq:hessian_omega}
 {\cal H}_{ij} = \frac{\partial^2{\bar \Omega}}{\partial X_i\partial X_j},
\end{equation}
with $(X_i=\sigma_x, \sigma_y, \alpha, \beta)$ evaluated at the points
where (\ref{eq:zero_gradient_condition}) is satisfied.

{Hence}, we must calculate the first and second derivatives of the
effective potential ${{\bar \Omega}}$ with respect to the order parameters in
order to find its minima.

Moreover, once we are interested in phase transitions, our algorithm
to find the minima considers the possibility of multiple minima. It
performs the following basic steps. With a given starting position in
the four-dimensional space of order-parameters, the algorithm looks
for a point with vanishing gradient
(\ref{eq:zero_gradient_condition}). As a second step, the eigenvalues
of the Hessian matrix (\ref{eq:hessian_omega}) are evaluated at this
point and their signs are checked: if they are all positive, this
point is a minimum and it is saved as such; otherwise it is either a
saddle-point or a maximum and, therefore, it is discarded. Next,
another starting position in the order-parameter space is tried, and
the procedure is repeated. If the point found is a minimum, it is
compared to the previous points: if all its four coordinates are 
sufficiently close to any of the other points (within $1\%$), it is 
discarded; otherwise, it is saved as a new minimum.
By sweeping a sufficiently
wide window in the parameter space, this is a simple algorithm able to 
find all the minima of the effective potential.

The thermodynamical state of equilibrium for each given $T$ and $\mu$
is given by the values of the order parameters at the global minimum
of the effective potential, found after the prescription described
above. This allows us to describe any thermodynamical quantity of
equilibrium. Particularly, we can study the behavior of the order
parameters as functions of the temperature and chemical potential, as
well as the phase diagram of the model. Moreover, finding the minima
of the effective potential at the vicinity of a first-order phase
transition allows the description of the dynamics of the
transition. More specifically, one can calculate the surface tension
between the two phases, as we discuss in the next section. Our results
will be discussed in Sec.~\ref{sec4}.

\subsection{The Saddle-Point Approach}

In the literature dedicated to the thermodynamics of strongly
interacting systems described by chiral models coupled to the
Polyakov loop, such as the PNJL and PQM models, the most common method
to determine the state of thermodynamical equilibrium is sometimes
called the {\it saddle-point} approach. It consists of finding the
points in which the in-medium effective potential of the theory has
vanishing derivatives with respect to the order parameters, as in
Eq.~(\ref{eq:zero_gradient_condition}). These extrema, however, are
not usually required to be minima of the effective
potential. Actually, the effective potential often has no minimum
whatsoever as a consequence of one of the two following reasons. 
Either the effective potential is
considered in full, so that it is a complex function of complex
variables (as we have discussed previously), or the Polyakov loop
variables are restricted to being real (instead of complex) numbers
(even before their expectation values are evaluated). In the first
case, no minimum can be defined because the effective potential is
complex valued. In the second case, it can be easily seen that the
effective potential is unbounded from below\footnote{Just consider
$\Phi=0$ in Eq.~(\ref{eq:Polyakov_potential_polynomial}) or in Eq.~(\ref{eq:Polyakov_potential_logarithmic}) 
and we can realize
that the Polyakov loop potential is unbounded from below for
$\bar\Phi\rightarrow\infty$.} \cite{Schaefer_07}.

The approach we propose in this section leads to predictions that
differ from those made by the saddle-point approach at
$\mu\not=0$. Some of these differences will be briefly discussed later 
on in Sec.~\ref{sec4}.

\section{\label{sec3}Nucleation in the PQM Model}

\subsection{Homogeneous Thermal Nucleation}

The dynamics of a first-order phase transition at small metastability
can be described by phenomenological droplet nucleation
models~\cite{Langer_73, Langer:1969bc,Gunton_83,Cahn_58,Cahn_59}. 
In such a family of models,
the transition between a metastable and a stable phase takes place by
the appearance and growth of domains (droplets or bubbles) of the
stable phase inside the metastable phase. The phase conversion is
finished when these domains grow and coalesce completely. In any case,
a minimum-sized bubble is needed for the beginning of the phase
transition, as can be inferred from the following heuristic
argument. The bulk free energy density of the metastable phase (often
called {\it false vacuum}) is, by definition, higher than that of the
stable phase (the {\it true vacuum}). Therefore, the conversion of a
given fraction of the system into the stable phase makes the bulk
free energy of the whole system lower. However, given that such a
conversion takes place within a connected domain of the system (most
likely in a spherical bubble~\cite{Coleman_78}), an interface is
needed in order to separate the (stable) interior from the
(metastable) exterior of this domain. Once the creation of an
interface represents an energy cost, the mechanism of phase
conversion through bubble nucleation settles a competition between the
free energy gain from the phase conversion of the bulk and the energy
cost from the creation of an interface. Roughly, one can say that the
free energy {\it shift} due to the appearance of a spherical bubble of the
stable phase of radius $R$ inside a metastable system is
\begin{eqnarray}
\label{eq:thin-wall_DeltaF}
 \Delta F_b &=& \left[\frac{4\pi}{3}R^3f_{\mrm{stable}} + 4\pi R^2
   f_{\mrm{wall}}\right] - \left[\frac{4\pi}{3}R^3 f_{\mrm{metastable}}\right]
 \cr\cr &=& \frac{4\pi}{3}R^3\Delta f + 4\pi R^2 f_{\mrm{wall}},
\end{eqnarray}
where $f_{\mrm{stable}}$ and $f_{\mrm{metastable}}$ are, respectively, the bulk
free energy densities of the stable and metastable phases, and
$f_{\mrm{wall}}$ is the surface energy density of the bubble wall, that is,
the {\it surface tension} of the interface between the two
phases. This formula clearly shows the competition between bulk
(negative) and surface (positive) contributions. Notice that the shift
in the bulk free energy 
($\Delta f \equiv f_{\mrm{stable}}-f_{\mrm{metastable}}<0$) is
proportional to the volume of the bubble, while the surface free
energy cost is proportional to its area. For the nucleation of small
bubbles, the energy cost is higher than the energy gain. Therefore,
small bubbles shrink. On the other hand, a very large bubble
represents a large bulk energy gain, which is higher than the surface
energy cost in creating the bubble. As a consequence, large bubbles
tend to grow even more and to occupy the whole system, completing the
phase transition. Consequently, this energy competition implies the
existence of a so-called {\it critical bubble}: any bubble smaller
than the critical bubble will shrink and any larger bubble will grow
and drive the phase conversion. For this reason, the critical bubble
is the crucial object in the theory of dynamical first-order phase
transitions of slightly metastable systems.

The appearance of a bubble (critical or not) of the stable phase
inside a metastable system is a natural consequence of the
never-ending thermal and quantum fluctuations of any thermodynamical
system sufficiently close to a first-order phase transition. As we
just discussed, each bubble created by these fluctuations may grow or
shrink, depending on its energy budget with regard to a homogeneous
metastable phase. One should also have in mind that larger
fluctuations (like a critical bubble) should be less common than
smaller ones. Although small bubbles are frequently created, they
rapidly disappear and do not contribute to the process of phase
conversion (with the exception in a weak first-order phase transition,
when coalescing subcritical
fluctuations~\cite{Gleiser:1992ed,Ramos:1996at} can complete the phase
transition without the nucleation of critical bubbles). Only those
that have a size equal to or larger than the critical bubble have a
decisive role. The smallest (and therefore the most probable) among
them is the critical bubble. This means that the mean time that it
takes for random fluctuations to create a critical bubble is the
shortest time scale for the creation of a lasting domain of the
stable phase, which is the dynamical seed of the phase conversion. 

Let us assume that the system is in a metastable state in
quasi-equilibrium with a reservoir with intensive coordinates
generically represented by ${\cal R}$ (e.g., temperature, chemical
potential, etc). Being metastable, bubbles of the stable phase with
different sizes randomly appear and subsequently disappear. This
process keeps happening until a critical bubble is nucleated and the
phase conversion effectively starts. It can be shown by different
approaches that the rate at which critical bubbles are nucleated per
unit time, per unit volume can be expressed in the form
\cite{Langer_73, Csernai:1992tj, Langer:1969bc,Coleman_77,Coleman_77_erratum,Callan_77}
\begin{equation}\label{eq:nucleation_rate}
 \Gamma({\cal R}) = {\cal P}({\cal R})\exp\left[-\frac{\Delta
     F_b({\cal R})}T\right],
\end{equation}
where $T$ is the temperature of the system in equilibrium with the
reservoir. The pre-exponential factor (or {\it prefactor}) ${\cal
  P}({\cal R})$ corresponds to the probability for a critical
bubble-like field fluctuation $\phi_b$ to be generated and grow
\cite{Langer_73,Csernai:1992tj,Gleiser:1993hf}. The last factor in
(\ref{eq:nucleation_rate}) is a Boltzmann factor in which $\Delta
F_b({\cal R})$ is the shift of free energy (as compared to the
homogeneous metastable phase) due to the formation of a critical
bubble. It can be easily shown that $\Delta F_b({\cal R})$ can be cast
as in (\ref{eq:thin-wall_DeltaF}) for a small degree of
metastability, where the thin-wall approximation is valid
\cite{Coleman_Book_88}
and where it is proportional to $f_{\mrm{wall}}^3/(\Delta f)^2$. 
In spite of the general importance of the prefactor in
(\ref{eq:nucleation_rate}), its specific form is not crucial for the
nucleation rate at a small degree of metastability.
Close to the coexistence of both phases, it shows a proportionality to 
$f_{\mrm{wall}}^{7/2}/\Delta f$, so that the nucleation rate
is strongly dominated by the exponential
factor \cite{Csernai:1992tj}. For this reason, it will be enough to focus 
on the free energy shift in this work.

The process of bubble nucleation in an impurity-free environment is
called {\it homogeneous nucleation}. In this work we shall only
consider the process of homogeneous nucleation, which is not the most
common in natural environments, such as in a boiling liquid. In such
cases, the presence of impurities can drastically accelerate the
nucleation of bubbles, and the process is called {\it inhomogeneous
 nucleation} (which is also the case when subcritical thermal
fluctuations can dominate~\cite{Gleiser:1992ed,Ramos:1996at}). The
process of inhomogeneous nucleation can be orders of magnitude faster
than homogeneous nucleation because impurities (like dust) often
reduce the free energy cost for the formation of a critical bubble,
raising the probability for its formation \cite{Kashchiev_Book_00,
Vehkamaeki_Book_06}. We do not consider it in this work for two
reasons. First, we wish to underestimate the nucleation rate and
inhomogeneities could only increase this rate, and second, we wish to keep the
approach as simple as possible.

\subsection{The Coarse-Grained Free Energy for a Single Scalar Order-Parameter}

As discussed in \cite{Linde_83,Coleman_Book_88, Weinberg_93,Gleiser:1993hf}, the
nucleation rate of critical bubbles can be calculated from the
microphysics using semiclassical methods in Euclidean thermal field
theory. We stress the importance of considering the effective action
in the problem of bubble nucleation, and not simply the effective
potential, once a critical bubble is clearly a non-homogeneous field
configuration. For simplicity, let us start with the Euclidean
Lagrangian density for a single scalar order-parameter field (which we
generically call $\phi$) of the form,
\begin{equation}\label{eq:scalar_lagrangian}
 {\cal L}_E = \frac12(\partial_\mu\phi)^2 + V(\phi).
\end{equation}
In this simple model, one assumes that the order parameter for a
system in thermodynamical equilibrium is given by the expectation
value of $\phi$. In general, it depends on the properties of the
reservoir, such as its temperature or chemical potential, which we
generally denote by ${\cal R}$. The (Euclidean) action is
\begin{equation}
 S_E[\phi,{\cal R}] = \int_0^\beta d\tau\int d^3x\,{\cal
   L}_E[\phi({\bf x},\tau)].
\end{equation}

In the high-temperature limit, $\beta\equiv1/T\rightarrow0$, the
imaginary time dependence of the order parameter can be neglected
\cite{Linde_83} and, therefore, we make the approximation
\begin{equation}
 S_E[\phi,{\cal R}]   \equiv  \frac{F[\phi,{\cal R}]}{T},
\end{equation}
where one identifies
\begin{equation}\label{eq:coarse-grained_free_energy}
 F[\phi,{\cal R}] = \int d^3x\,\left[\frac12(\nabla\phi)^2 +
   V_{\mrm{eff}}(\phi,{\cal R})\right]\;,
\end{equation}
with the {\it coarse-grained free energy} of the system. Notice that
the coarse-grained free energy is a sort of dimensionally reduced
effective action, so that the tree-level potential $V(\phi)$ must be
replaced by the medium-dependent effective potential
$V_{\mrm{eff}}(\phi,{\cal R})$. In full thermodynamical equilibrium, the
minimization of the coarse-grained free energy (which is equivalent
to the minimization of the Euclidean action) is achieved by a constant
field configuration $\phi({\bf x})=\phi_0$ so that the gradient term
vanishes and $V_{\mrm{eff}}(\phi_0,{\cal R})$ must be a global minimum of
$V_{\mrm{eff}}$.

The possibility of a metastable state arises when $V_{\mrm{eff}}$ develops
some local minimum other than the global minimum at $\phi=\phi_0$. In
this framework, a metastable state is described by a constant field
configuration $\phi_f$ that is a {\it local minimum} of
$V_{\mrm{eff}}$. For this reason, this second minimum is often called a
{\it false vacuum} of the potential, while the global minimum is
called the {\it true vacuum} of the theory, $\phi_t\equiv \phi_0$. 
A bubble is then
represented as a non-homogeneous spherically symmetric field
configuration $\phi(r)$ such that~\cite{Scavenius_01,Bessa_09,Coleman_Book_88}
\begin{equation}\label{eq:bubble_bc}
 \lim_{r\rightarrow\infty}\phi(r)=\phi_f\;\;\;\;\;\mbox{and}\;\;\;\;\;\frac{d\phi}{dr}(0)=0,
\end{equation}
where $\phi_f$ is the value of the order-parameter field at the false
vacuum. That is, away from the center of the bubble, the system is in
the metastable phase. But, in the vicinity of its center, the field
configuration should be close to the stable minimum (but not
necessarily exactly on it).

The critical bubble is a saddle point field configuration $\phi_b$
that extremizes the functional $F$, {\it i.e.}, it solves the Euler-Lagrange
equation
\begin{equation}\label{eq:general_bubble_eq}
 \frac{\delta F[\phi;\, T]}{\delta\phi({\bf x})}=0\;\;\Rightarrow\;\;
 \nabla^2\phi({\bf x}) - \frac{\partial
   V_{\mrm{eff}}}{\partial\phi}[\phi({\bf x})] = 0.
\end{equation}

It can be shown \cite{Coleman_78} for a wide class of Lagrangians,
including (\ref{eq:scalar_lagrangian}), that the smallest value of
$F$ indeed corresponds to a spherically symmetric solution of
(\ref{eq:general_bubble_eq}), $\phi(r)$, so that the equation to be
solved is now the non-linear ordinary differential equation
\begin{equation}\label{eq:spherical_bub_eq}
 \frac{d^2\phi(r)}{dr^2} + \frac{2}{r}\frac{d\phi(r)}{dr} =
 \frac{\partial V_{\mrm{eff}}}{\partial\phi}[\phi(r)],
\end{equation}
with the boundary conditions (\ref{eq:bubble_bc}).

The coarse-grained free energy associated with a spherical bubble
$\phi_b$ is then
\begin{equation}\label{eq:spherical_deltaF}
F_b = 4\pi\int_0^\infty dr
 \,r^2\left\{\frac{1}{2}\left[\frac{d\phi_b(r)}{dr}\right]^2 +
 V_{\mrm{eff}}[\phi_b(r)] \right\},
\end{equation}
which directly follows from (\ref{eq:coarse-grained_free_energy}).

Once the solution for (\ref{eq:spherical_bub_eq}) and
(\ref{eq:bubble_bc}) is found, that is, the critical bubble profile
$\phi_\mrm{crit}(r)$, the shift in the coarse-grained free energy due
to the appearance of a critical bubble,
\begin{equation}\label{eq:free_energy_shift}
 \Delta F_b({\cal R}) = F[\phi_\mrm{crit};\, {\cal R}] - F[\phi_f;\, {\cal R}],
\end{equation}
needed in the nucleation rate (\ref{eq:nucleation_rate}), can be
readily calculated.

For a generic effective potential $V_{\mrm{eff}}$, the solution of
(\ref{eq:spherical_bub_eq}) with boundary conditions
(\ref{eq:bubble_bc}) cannot be obtained analytically. However, an
approximate solution can be found when the system is very close to
the coexistence line, so that it is slightly metastable and the {\it
 thin-wall approximation} \cite{Coleman_77, Coleman_Book_88, Linde_83} 
is applicable. Within these limits, the coarse-grained free energy shift
(\ref{eq:free_energy_shift}) can be well approximated by the
expression
\begin{equation}\label{eq:free_energy_shift_thin-wall}
 \Delta F_b({\cal R}) = \frac{16\pi}{3}\frac{\Sigma^3}{(\Delta
   V_{\mrm{eff}})^2},
\end{equation}
where
\begin{equation}\label{eq:sigma_definition}
 \Sigma(T)\equiv\int_0^\infty\,dr\left[\frac{d\phi_\mrm{crit}(r)}{dr}\right]^2\;,
\end{equation}
is the {\it surface tension} of the critical bubble interface between
the phases. Notice that the surface tension is calculated directly
from the critical bubble solution $\phi_{crit}$. It must be so because the
surface tension must contain information about how the system reacts
to inhomogeneities (e.g., a wall). That is, any description of a
critical bubble has to take into account more than just the bulk
thermodynamics. This is the reason why the coarse-grained free energy
(\ref{eq:coarse-grained_free_energy}) is needed in the formalism for
bubble nucleation.

The quantity $\Delta V_{\mrm{eff}}=V_{\mrm{eff}}(\phi_t) - 
V_{\mrm{eff}}(\phi_f)$
is the difference between the bulk free energy in the two homogeneous
vacua. In the grand canonical potential, it can be identified as
$\Delta V_{\mrm{eff}}(T,\mu) \equiv - \Delta p(T,\mu)$, {\it i.e.}, minus the
difference of pressures between the two phases. Using the thin-wall
approximation, the surface tension integral
(\ref{eq:sigma_definition}) can be calculated without solving for the
profile. After changing variables from $r$ to $\phi$, one finds
\begin{equation}\label{eq:sigma_thinwall}
 \Sigma({\cal R}) =
 \int_{\phi_t}^{\phi_f}d\phi\,\sqrt{2V_{\mrm{eff}}[\phi;\,{\cal R}]},
\end{equation}
so that only the effective potential $V_{\mrm{eff}}[\phi;\,{\cal R}]$ is
needed to calculate $\Sigma({\cal R})$ in the thin-wall
approximation. Notice that $V_{\mrm{eff}}$ is normalized so that its global
minimum is located at $\phi=\phi_t=0$. Exactly at the coexistence points,
where the thin-wall approximation is exact, the minima are degenerate
with $V_{\mrm{eff}}=0$.

As a last remark, let us notice that the surface tension cannot be
correctly defined unless $\phi_t$ and $\phi_f$ are actual minima of
the effective potential. This a further motivation for the careful
approach discussed in the previous Sec.~\ref{sec2}.


\subsection{The Coarse-Grained Free Energy for the Polyakov-quark-meson Model}

In order to calculate the free energy shift $\Delta F_b({\cal R})$ due
to the nucleation of a critical bubble within the PQM model, we first
need to define the coarse-grained free energy functional and then
identify an order parameter.

Just as in the case of a single order-parameter, the coarse-grained
free energy of the PQM model has its origin in the in-medium
effective action of the theory. In the PQM model, the Lagrangian 
density of the chiral fields $\sigma_x$ and $\sigma_y$ directly leads to a
kinetic term of the form $|\nabla\phi|^2$. The kinetic term for
the Polyakov loop variable, however, is not determined {\it a
priori}. For simplicity, once we consider the Polyakov loop order
parameters $\alpha$ and $\beta$ as independent, real variables at
finite $\mu$, we assume the kinetic term
\begin{equation}\label{eq:polyakov-loop_kinetic-term}
 {\cal L}_\mrm{kin}[\alpha,\beta] = \frac{\kappa^2}2
 (\partial_\mu\alpha)^2 + \frac{\kappa^2}2(\partial_\mu\beta)^2,
\end{equation}
which is motivated by the $Z(3)$-line model of
Ref.~\cite{Pisarski_00}.

Effective models cannot do much more than estimate the value of the
kinetic parameter $\kappa$. From dimensional arguments, it is
estimated to be $\kappa^2=N_cT_0^2/g_s^2$ \cite{Pisarski_00,Scavenius_02}. 
Assuming $N_c=3$ and $\alpha_s=g_s^2/4\pi \simeq0.3$, we find 
$\kappa\simeq0.9T_0$, which is of the order of magnitude of the only 
scale in the pure glue model, the transition temperature $T_0$. Another 
consistent approach is to consider as an input the surface tension of 
the pure gauge $SU(3)$ theory calculated through lattice Monte-Carlo 
simulations \cite{Beinlich_97,Lucini_05}and in an effective matrix model \cite{Dumitru_12},
$\Sigma_{SU(3)}\simeq0.016 T_0^3$.
With a given parametrization of the Polyakov loop
potential, the parameter $\kappa$ can be fitted from the value of the
surface tension at $T=T_0$ using Eq.~(\ref{eq:sigma_thinwall}). We
adopt this second approach here in order to fix the parameter
$\kappa$ with the implications discussed in Sec.~\ref{sec4}.

The coarse-grained free energy for the PQM model can be written as
\begin{multline}\label{eq:PQM_coarse-grained_free-energy}
 {\cal F}[\sigma_x, \sigma_y, \alpha, \beta]=\int d^3x
 \left[\frac12(\nabla\sigma_x)^2 + \frac12(\nabla\sigma_y)^2
   +\right.\\ \left.+ \,\frac{\kappa^2}2(\nabla\alpha)^2
   +\frac{\kappa^2}2(\nabla\beta)^2 + {\bar \Omega}(\sigma_x, \sigma_y,
   \alpha, \beta)\right],
\end{multline}
where ${\bar \Omega}(\sigma_x, \sigma_y, \alpha, \beta)$ is the effective 
potential defined in Eq.~(\ref{eq:effective_potential_alphabeta}).

As discussed previously, the critical bubble is a solution of the {\it
four} coupled Euler-Lagrange equations that arise from the
functional (\ref{eq:PQM_coarse-grained_free-energy}). There is no
general procedure to solve this set of coupled equations, even though
some {\it ansatz} solutions can be eventually tried for very simple
effective potentials \cite{Rajaraman_Book_87}.

There are two possible ways in which we can tackle the problem. The
first is, of course, to numerically solve the four equations that
follow from the extremization of ${\cal F}$ simultaneously. 
The exact solution will define a
path in the four-dimensional space of order parameters. Notice,
however, that this path is in general not in the ``valley'' that
connects the two minima, as can be intuitively seen from the inverted
potential mechanical analog.

Let us now argue that our approach will lead us to an {\it overestimate} 
of the surface tension or, equivalently, to an underestimate of the 
nucleation rate. According to Eq.~(\ref{eq:nucleation_rate}), the higher 
the free energy shift of the
critical bubble, the lower is the nucleation rate (at least as long
as the thin-wall approximation is valid). Now consider that the exact
critical bubble solution is found. Consider also a small deviation
from it that follows a different path in the order-parameter
space. Once the true critical bubble is a saddle point solution in a
functional space, one expects that the distorted path will have a
{\it higher} value of $\Delta F$ as compared with the true bubble. Hence,
we can (artificially) constrain the configuration path to a given
arbitrary line that connects the two vacua in the space of order
parameters. This path will give us an {\it overestimate} of the
free energy and, therefore, of the surface tension. The simplest
choice is, of course, a straight line that connects both
minima.\footnote{In the whole range of interest in the $T-\mu$ plane,
we have always found either one or two minima. The case of three or
more minima, which we have not encountered, would lead to a much
more complicated analysis.} For example, let $\sigma_x^{(1)}$
and $\sigma_x^{(2)}$ be the values of the $\sigma_x$ order-parameter
in the two minima of ${\bar \Omega}$ close to the coexistence line. The
interpolation
\begin{equation}\label{eq:straight_line_ansatz}
 \sigma_x = \xi \sigma_x^{(1)} + (1-\xi)\sigma_x^{(2)}
\end{equation}
is such that for $0\leq\xi\leq 1$, the value of $\sigma_x$ varies
from one minimum to the other. If the same function $\xi$ is used to
parametrize the path followed by the remaining order parameters, 
\begin{eqnarray}
  \sigma_y = \xi \sigma_y^{(1)} +
  (1-\xi)\sigma_y^{(2)},\nonumber\\ 
\alpha = \xi \alpha^{(1)} +
  (1-\xi)\alpha^{(2)},\nonumber\\ 
\beta = \xi \beta^{(1)} +
  (1-\xi)\beta^{(2)},
\end{eqnarray}
then this path is a straight line in the four-dimensional order
parameter space.

It is now natural to define the four-dimensional order-parameter for
the PQM model as
\begin{equation}\label{eq:PQM_order-parameter}
 {\vec \Psi} = \left(\sigma_x, \sigma_y, \kappa\alpha,
 \kappa\beta\right).
\end{equation}

It can be easily shown from Eqs.~(\ref{eq:straight_line_ansatz}) and
(\ref{eq:PQM_order-parameter}) that if we write the coarse-grained
free energy (\ref{eq:PQM_coarse-grained_free-energy}) in terms of the
field $\xi$, it assumes the form 
\begin{equation}\label{eq:coarse_grained_free_energy_xi}
 \tilde{\cal F}(\xi) = \int
 d^3x\left[\frac{h^2}2\left(\nabla{\xi}\right)^2 + \tilde{\Omega}(\xi)
   \right],
\end{equation}
where 
\begin{eqnarray}\label{eq:definition_h2}
 h^2&=& \left(\vec\Psi_0 - \vec\Psi_1\right)^2 \crcr 
&=&(\Delta\sigma_x)^2 + (\Delta\sigma_y)^2 +
 (\kappa\Delta\alpha)^2+(\kappa\Delta\beta)^2,
\end{eqnarray}
and $\tilde{ \Omega}(\xi)$ is the projection of the effective potential
${\bar \Omega}(\vec\Psi)$ along the straight line defined by
(\ref{eq:straight_line_ansatz}). Notice that the coarse-grained free
energy $\tilde{\cal F}$ is formally equivalent to the single field
coarse-grained free energy (\ref{eq:coarse-grained_free_energy}) and
we can now consider $\xi$ as a scalar order-parameter. As a result,
the surface tension in the PQM model can be overestimated by
\begin{equation}\label{eq:sigma_thinwall_straight_line}
 \Sigma({\cal R}) = h\int_0^1d\xi\,\sqrt{2\tilde{ \Omega}[\xi;\,{\cal
       R}]},
\end{equation}
where the domain of integration ranges from one minimum of ${\bar \Omega}$
(and thus of $\tilde{\Omega}$) to the other along a straight line path
in the four-dimensional order-parameter space.

Notice that a path other than a straight line in the space of order
parameters would not lead to a simple coarse-grained free energy with
the same form as (\ref{eq:coarse_grained_free_energy_xi}). In the
general case, even the definition of the surface tension should be
reformulated. This problem will be treated in a separate work.

In the following section, we present our results for the equilibrium
thermodynamics as well as for the surface tension in the PQM model.

\section{\label{sec4}Results and Discussion}

We will now present our original results for the $N_f=2+1$ PQM model
for quantities in strict thermodynamical equilibrium, as well as an
overestimate of the surface tension of a hadron-quark interface as
described by this model. We use the methods and approximations
discussed in the previous sections. Here, we call the Polynomial
parametrization of \cite{Scavenius_02} ``Poly-I'', while that of
Ref.~\cite{Ratti_06} is called ``Poly-II''. The logarithmic
parametrization of \cite{Roessner_07} is referred to as ``Log''.

Notice that the imaginary part of the effective potential
(\ref{eq:imaginary_part_omega_qqbar}) is zero for $\mu=0$, and our
scheme has to be completely equivalent to the {\it saddle-point}
approach in this limit (notice that $\alpha\equiv\Phi$ for
$\mu=0$). Our numerical results show that this is true indeed. 

In an explicit crosscheck, we calculated the pseudo-critical
temperatures for deconfinement and chiral symmetry restoration at
$\mu=0$, with the same parameter set of Ref.~\cite{Schaefer_10}. The
chiral pseudocritical temperature is defined as the peak in the
chiral susceptibility $\chi_x \equiv
-\partial\langle\sigma_x\rangle/\partial T$, while the deconfinement
pseudocritical temperature takes place at the peak of $\chi_P \equiv
\partial\langle\alpha\rangle/\partial T$. We found the same
pseudocritical temperatures as those in \cite{Schaefer_10} for the
Poly-I and the logarithmic parametrizations at $\mu=0$. This shows
numerically a consistency between our approach and the saddle-point
approximation in the regime where both methods have to lead to the
same results\footnote{We refer the interested reader to
Ref.~\cite{Schaefer_10} for the actual values of pseudo-critical
temperature found within the $N_f=2+1$ PQM model.}. However, we
must stress, the two approaches do give different predictions at
finite quark chemical potential, as expected. This difference is
mainly due to the fact that we are explicitly considering only the
real part of the effective potential at finite chemical potential. As
we have already discussed in Secs.~\ref{sec1} and \ref{sec2}, this is
a necessary condition in order to fulfill the need of defining minima and 
for their
existence in the thermodynamical potential, which can only
be achieved through the reality of the effective potential. In the
following, we present our original results alongside the
corresponding calculations we performed using the saddle-point
approach for comparison.

\begin{figure}
\centering
\includegraphics[width=0.47\textwidth]{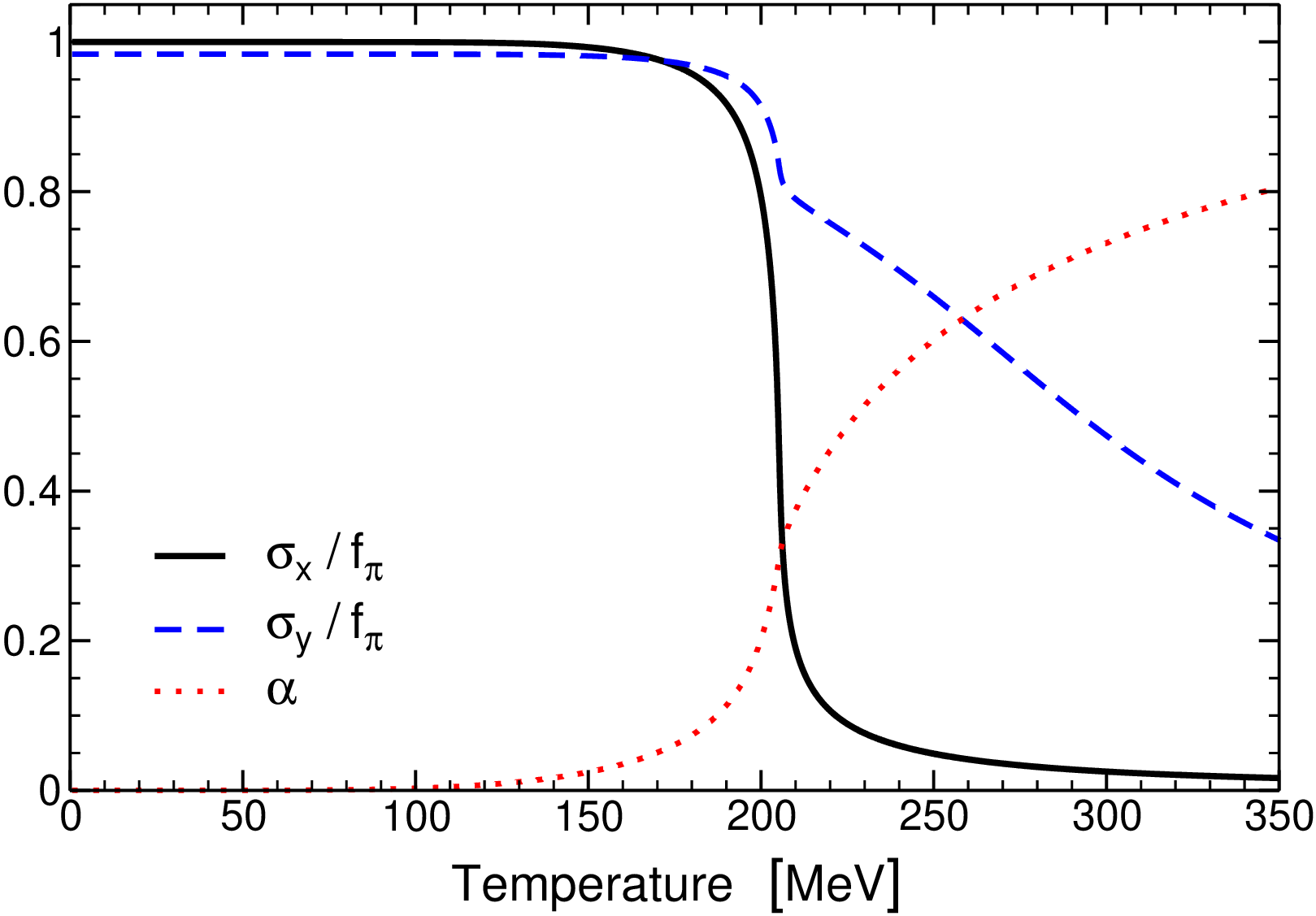}
\caption{(Color online.) Normalized nonstrange condensate
  ($\langle\sigma_x\rangle/f_\pi$, solid line) and strange condensate
  ($\langle\sigma_y\rangle/f_\pi$, dashed line) and Polyakov loop
  variable ($\langle\alpha\rangle$, dotted line) as a function of the
  temperature for $\mu=0$. The parameter set used is that of Table
  \ref{tab:chiral_pot_constants} with $m_\sigma=600~$MeV and
  $T_0=270\,$MeV together with the logarithmic Polyakov loop potential.}
\label{fig:polyakov_loop}
\end{figure}

Let us first discuss our results for the behavior of the order
parameters for the model at $\mu=0$ for the logarithmic
parametrization of the Polyakov loop potential. The behavior for the
polynomial parametrizations Poly-I and Poly-II are very similar and
we do not display them here. Our choice of parameters is shown in
Table~\ref{tab:chiral_pot_constants}, with $m_\sigma=600\,$MeV and
$T_0=270\,$MeV. In Fig.~\ref{fig:polyakov_loop}, we show our results
for the expectation value of the chiral condensates $\sigma_x$ and
$\sigma_y$ (normalized by the vacuum value of the nonstrange
condensate $\langle\sigma_x\rangle_{vac}=f_\pi$), as well as for the
Polyakov loop variable $\alpha$.
The steep change of the order
parameters at around $T=205\,$MeV indicates the pseudo-critical
temperature of both the chiral and the deconfinement crossovers. (For
some other choices of parameters, these pseudo-critical temperatures
may not coincide \cite{Schaefer_10}.)

The low-temperature behavior of $\langle\sigma_x\rangle$ and of
$\langle\sigma_y\rangle$ shows little sensitivity to the choice of the 
Polyakov
loop potential, being dominated by the chemical potential-dependent
contribution from the mesonic sector. 

\begin{figure}
\centering
\includegraphics[width=0.47\textwidth]{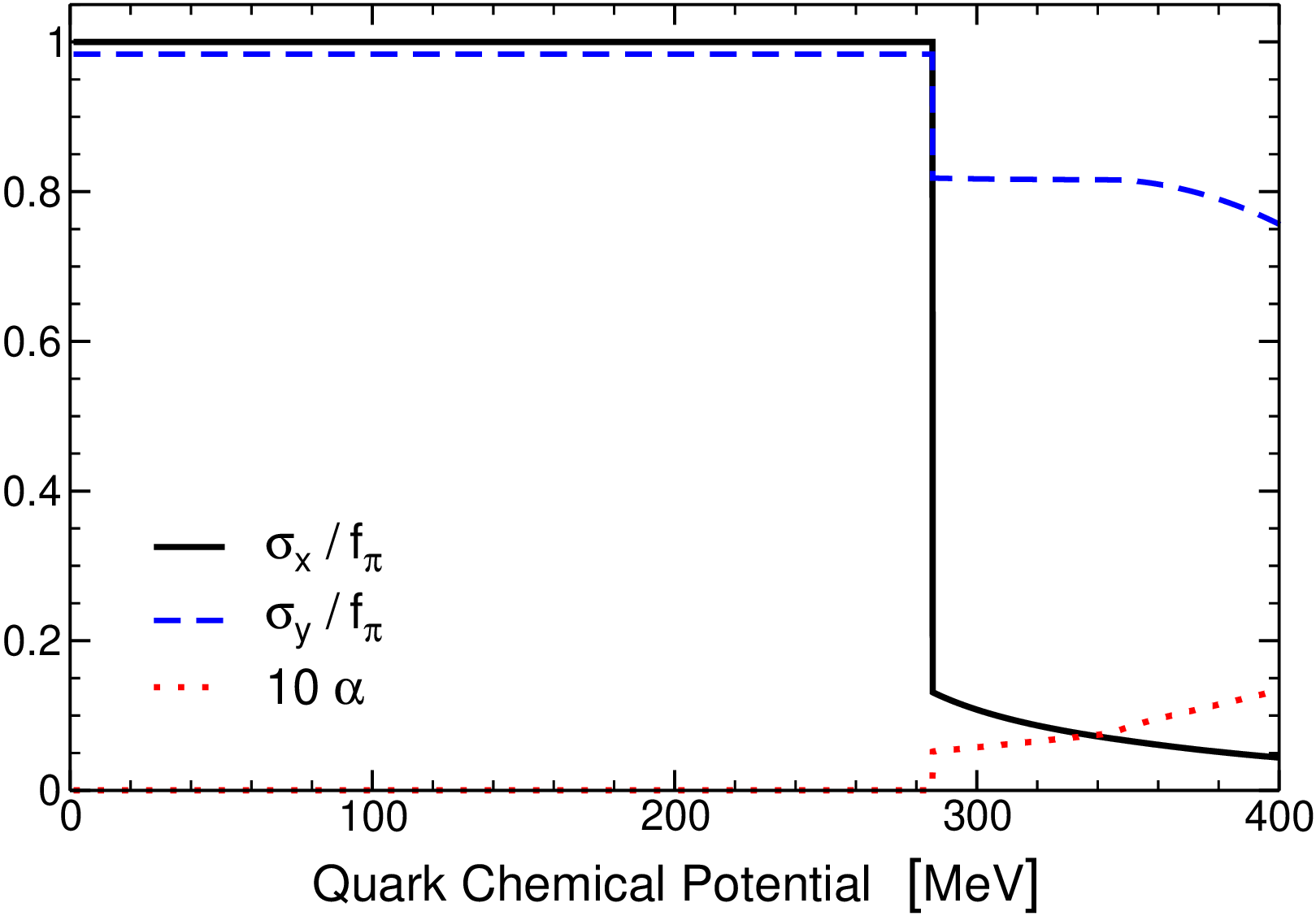}
\caption{(Color online.) Normalized nonstrange condensate
  ($\langle\sigma_x\rangle/f_\pi$, solid line) and strange condensate
  ($\langle\sigma_y\rangle/f_\pi$, dashed line) and Polyakov loop
  variable ($\langle\alpha\rangle$, dotted line) as a function of the
  chemical potential for $T=2\,$MeV. The parameter set used is that of
  Table \ref{tab:chiral_pot_constants} with $m_\sigma=500\,$MeV and
  $T_0=182\,$MeV. For the Polyakov loop potential the logarithmic
  parametrization is chosen but at this low temperature its influence is negligible.}
\label{fig:chiral_condensates_T0}
\end{figure}

Next, we observe that, at low temperatures, the order parameters have
a discontinuity at some finite $\mu$. This signals a first-order
phase transition. In Fig.~\ref{fig:chiral_condensates_T0} we show the
behavior of the condensates $\sigma_x$ and $\sigma_y$ and the Polyakov
loop variable $\alpha$ (multiplied by a factor of ten) as a function
of the chemical potential for a very low temperature $T=2\,$MeV with
sigma mass $m_\sigma=500\,$MeV. Notice that the Polyakov loop variable
has a very low expectation value, which grows very slowly at such a
low temperature, even at high densities. 

\begin{figure}
\centering
\includegraphics[width=0.47\textwidth]{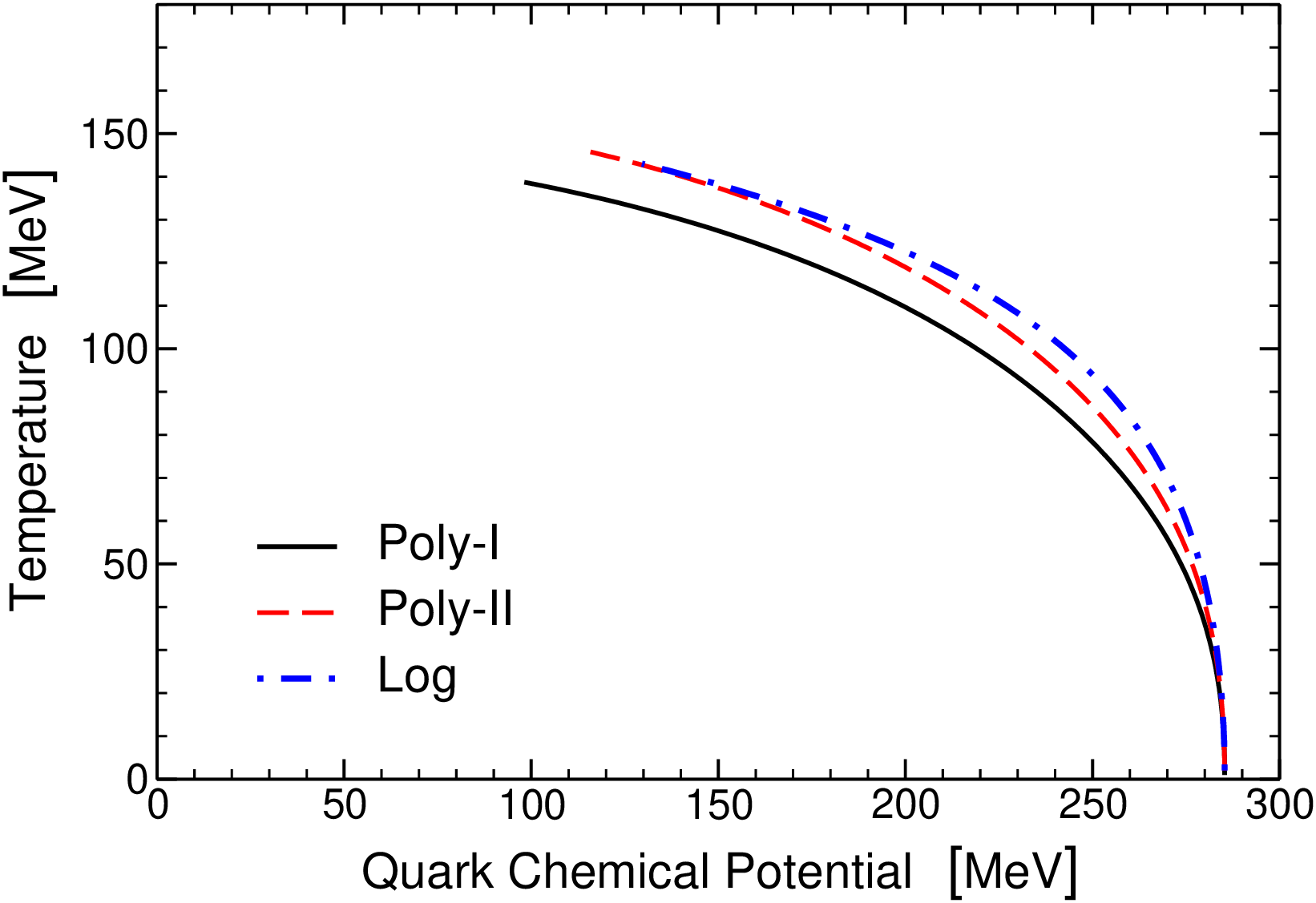}
\caption{(Color online.) Lines of first-order phase transition of the
  PQM model. The sigma meson mass is $m_\sigma=500\,$MeV and
  $T_0=182\,$MeV. Parametrizations of the Polyakov loop potential:
  Poly-I (full black line), Poly-II (dashed red line) and Log
  (dash-dotted blue line).}
\label{fig:phase_diagram}
\end{figure}

Having found a crossover at $\mu=0$ and a first-order transition at
$T=0$, the next step is to calculate the complete $T-\mu$ phase
diagram of the model. The phase diagram is shown in 
Fig.~\ref{fig:phase_diagram} for
$m_\sigma = 500\,$MeV and $T_0=182\,$MeV for the three parametrizations
of the Polyakov loop potential we have considered in this work.

As mentioned before, the saddle-point approach leads to different
results to quantities in thermodynamical equilibrium. Comparing the phase
diagrams obtained within the two methods, we notice that they are
equivalent at $\mu=0$ (where the sign problem does not exist) and at
$T=0$ (where only the chiral fields are relevant and, again, the sign
problem does not exist). For this reason, the phase diagrams can
only differ between these two extrema and, in particular, they can
lead to different positions for the critical end point
(CEP). Table~\ref{tab:CEP_coordinates} shows the $T-\mu$ coordinates
of the CEP found using the method we propose and the saddle-point
method.
Depending on the parametrization and value of $T_0$ the positions of
the CEP found within the two methods can differ
significantly. The difference increases with a larger value of the glue
transition temperature $T_0$ and is most pronounced if using the Poly-II
parametrization. Due to the slope of the phase transition line around
the CEP the differences of the critical chemical potential is larger
than that of the temperatures.

\begin{table}
  \caption{Temperature and quark chemical potential coordinates of the
    critical end point (CEP) found with the minimization
    procedure of Sec.~\ref{sec2}, $(T_C,\mu_C)$, and with the saddle-point
    approach, $(T_C,\mu_C)^S$. We choose a sigma meson mass
    $m_\sigma=500\,$MeV and pure glue transition temperatures
    $T_0=182\,$MeV (with effective screening) or $T_0=270\,$MeV (pure
    gauge). The parameters not shown in this table are found in Table
    \ref{tab:chiral_pot_constants}.}
  \begin{ruledtabular}
  \begin{tabular}{l|c|c||c}
    Parametrization			& $T_0$ [MeV]	& $(T_C,\mu_C)\!\!$ [MeV]	& $(T_C,\mu_C)^S\!\!$ [MeV]  \\\hline
    Log \cite{Roessner_07}		& 182         	& $(143,129)$             		& ${(143,128)}$ \\\hline
    						& 270         	& $(192,88)$    				& ${(192,84)}$ \\\hline
    Poly-I \cite{Scavenius_02}	& 182         	& $(139,99)$				& ${(140,92)}$ \\\hline
    						& 270        		& $(171,103)$				& ${(175,83)}$ \\\hline 
    Poly-II \cite{Ratti_06} 		& 182         	& $(146,115)$              		& ${(152,80)}$ \\\hline
    						& 270         	& $(176,129)$ 			         & ${(184,103)}$ \\
  \end{tabular} 
  \end{ruledtabular}
  \label{tab:CEP_coordinates}
\end{table}

\begin{figure}
\centering
\includegraphics[width=0.47\textwidth]{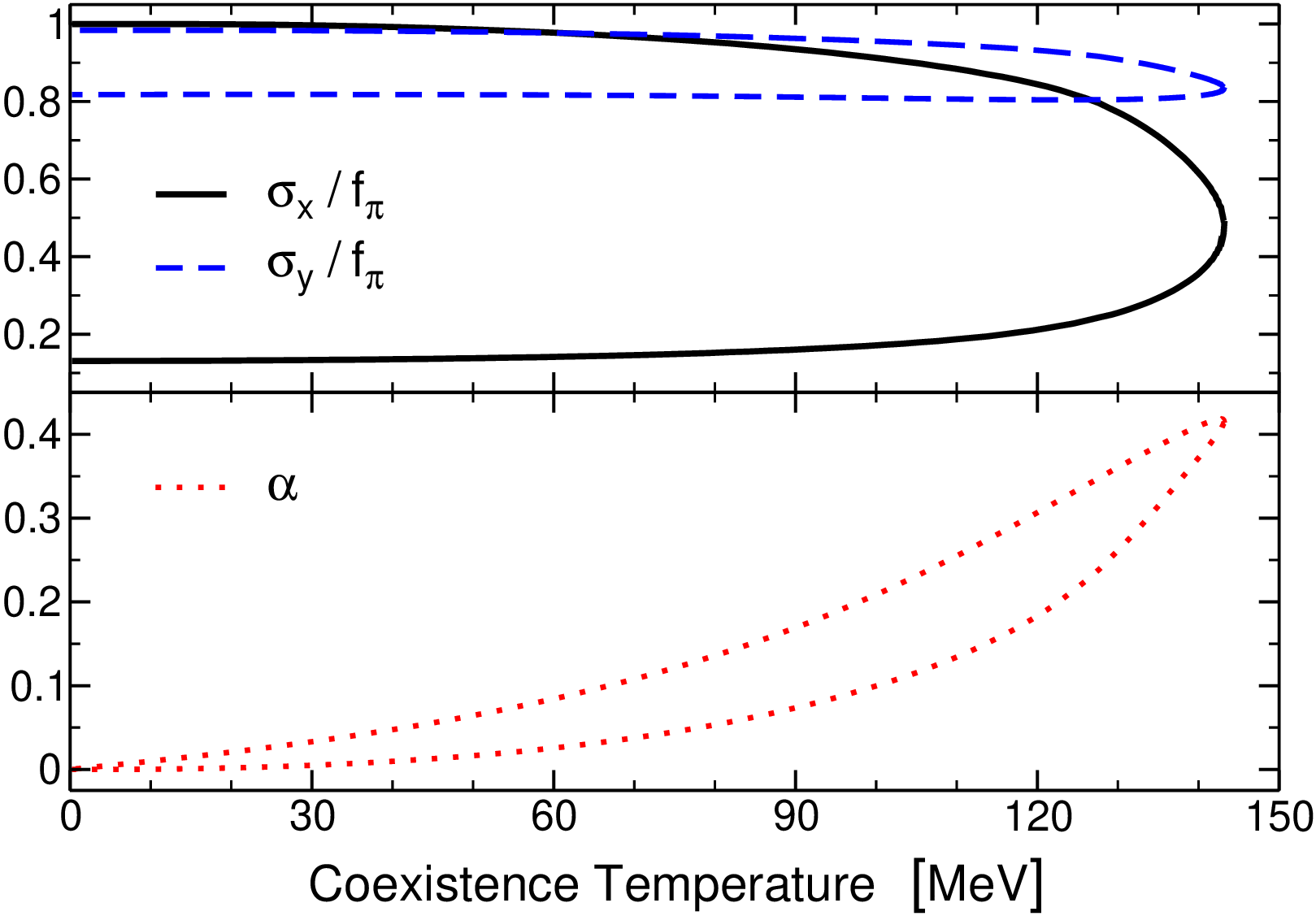}
\caption{(Color online.) Degenerate values of the chiral condensates
  $\sigma_x$ and $\sigma_y$ (upper part) and Polyakov loop variable 
  $\alpha$ (lower part) over the first-order line in the $T-\mu$ plane as a
  function of the coexistence temperature. Notice that the minima merge smoothly
  at the critical end point. The parameter set used is that of Table
  \ref{tab:chiral_pot_constants} with $m_\sigma=500\,$MeV and
  $T_0=182\,$MeV together with the logarithmic parametrization of the Polyakov loop potential.}
\label{fig:chiral_condensates_coexistence}
\end{figure}

The effective potential has always two degenerate minima over the
coexistence line of the phase diagram, each one with a different
quartet of expected order parameters. In
Fig.~\ref{fig:chiral_condensates_coexistence}, we show how the
coordinates of these minima evolve {\it along the first-order
line of the phase diagram}, starting from $T=0$ and finite $\mu$ up to the CEP. 
Notice that the two minima smoothly merge at the CEP with
diverging derivative with respect to the temperature. This signals
the expected second-order phase transition at the CEP.

\begin{figure}
\centering
\includegraphics[width=0.47\textwidth]{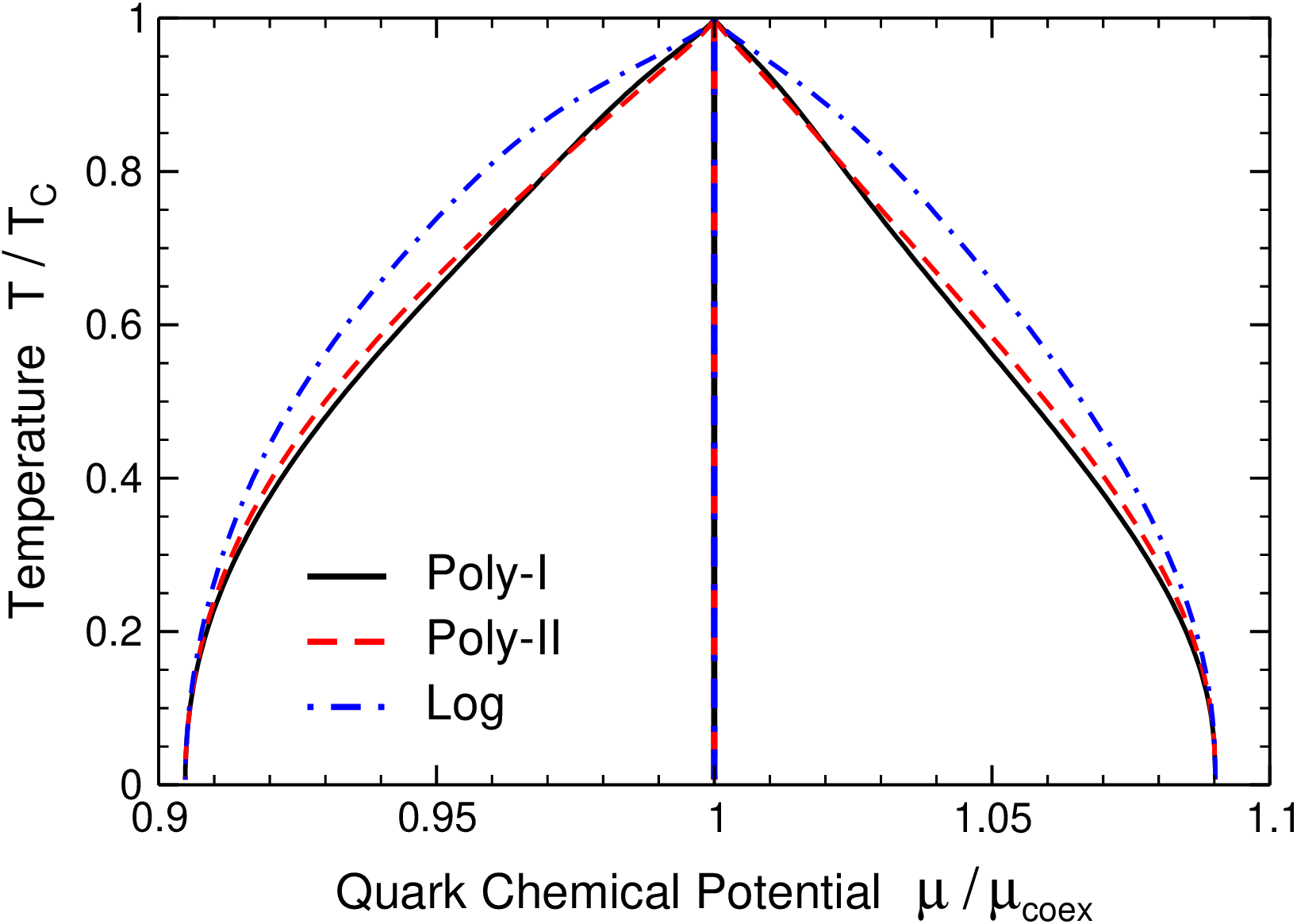}
\caption{(Color online.) Metastable regions of the
  PQM model for the three considered parametrizations. 
  We compare their extension with respect to the 
  coexistence line ($\mu_\mrm{coex}$). The temperature is given
  relative to the one of the critical end point $T_{C}$, 
  see Table \ref{tab:CEP_coordinates}.
  The sigma meson mass is $m_\sigma=500\,$MeV and
  $T_0=182\,$MeV.}
\label{fig:metastable_region}
\end{figure}

The two minima of the potential that are degenerate on the coexistence line 
persist as a global and a metastable local minimum in some region of the phase
diagram around the phase transition line. Going away from the coexistence line, 
the intervening maximum approaches the local minimum until these two extrema meet 
and form an inflection point that defines the spinodal line.
In Fig.~\ref{fig:metastable_region} we show the extension of the metastable region 
of the $T-\mu$ phase diagram that is limited by the spinodal lines. The extension 
is relatively independent of the parametrization of the Polyakov loop potential. 
We find that the degree of metastability that can be reached is relatively modest.

\begin{figure}
\centering
\includegraphics[width=0.47\textwidth]{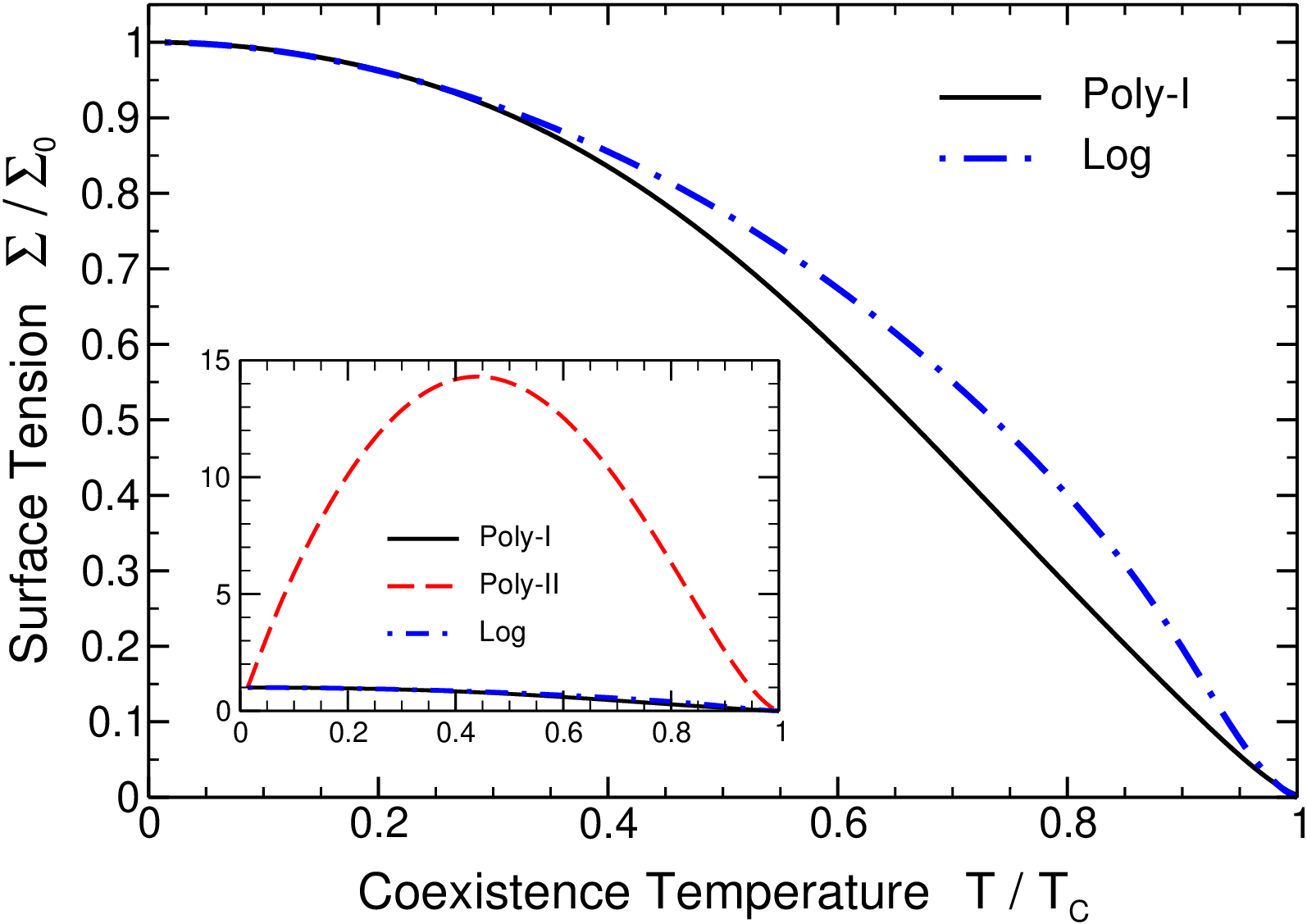}
\caption{(Color online.) Surface tension relative to its zero-temperature value along the first-order
  transition line in the $T-\mu$ phase diagram as a
  function of the temperature relative to the one of the critical end point. We use the Poly-I and Log
  parametrizations with $T_0=182$MeV. Inset: The Poly-II 
  parametrization is shown for comparison. The parameter set used is that of Table
  \ref{tab:chiral_pot_constants} with $m_\sigma=500\,$MeV. The other parameters can be found in Tables 
  \ref{tab:CEP_coordinates} and \ref{tab:zeroT_surfacetension}.}
\label{fig:surface_tension_PolyII}
\end{figure}

Let us now study the behavior of the surface tension in the PQM model
along the coexistence line for different
parametrizations of the Polyakov loop potential $\cal U$. The surface
tension was evaluated in the thin-wall approximation, according to
Eq.~(\ref{eq:sigma_thinwall}). 
In Fig.~\ref{fig:surface_tension_PolyII}, we show the behavior of the
surface tension along the first-order transition line for the Poly-I,
Poly-II and the Log parametrizations of the Polyakov loop
potential. The surface tension for the Poly-II potential is off scale
to the other two ones and, therefore, it is shown in the inset of
Fig.~\ref{fig:surface_tension_PolyII}.

Even though equilibrium observables like the behavior of the order
parameters and the phase diagram are quite similar for the three
parametrizations, there is a remarkable difference between them
regarding the surface tension, in particular for the result obtained for the
Poly-II parametrization. We interpret this as an artifact of the
Poly-II parametrization. Although all parametrizations lead to very
similar descriptions of the thermodynamics of the deconfinement
transition in equilibrium, they offer different behaviors for the
order parameters. At the pure glue transition temperature $T_0$, the
Polyakov loop potentials present two minima, which correspond to the
two coexisting values of $\Phi=\bar\Phi$ at $T=T_0$. One of these
minima is always at $\Phi=0$, representing the confined phase, while
the other sits at some nonzero value $\Phi=\Phi_0$. Using the
parameter sets shown in Table~\ref{tab:Ploop_pot_params}, one can
easily see that the Poly-I parametrization possesses the second
minimum at $\Phi_0\simeq0.6$, while $\Phi_0\simeq 0.07$ for the
Poly-II parametrization and $\Phi_0\simeq0.5$ for the Log
parametrization. A comparison to the pure gauge $SU(3)$-lattice
calculation of \cite{Kaczmarek_02,Mykkanen_12} shows that one has 
$\Phi_0^\mrm{latt}\simeq 0.4$. 

The reason why our results for the surface tension with the Poly-II
parametrization are so different can be understood from the
definition (\ref{eq:definition_h2}). The kinetic parameter $\kappa$ in
(\ref{eq:polyakov-loop_kinetic-term}) is fitted from the value of the
surface tension in pure gauge $SU(3)$-lattice simulations using
Eq.~(\ref{eq:sigma_thinwall}). After some simple manipulations, one
sees that the smaller the barrier between the minima of the pure gauge
potential, the larger $\kappa$ has to be in order to reproduce the
known value of $\Sigma_{SU(3)}\simeq0.016 T_0^3$ \cite{Beinlich_97,Lucini_05,Dumitru_12}.
Once the Poly-II parametrization
leads to a very short and narrow barrier, it gives the large value
$\kappa\simeq160\,T_0$. This is three orders of magnitude larger than the
values given by the Poly-I and Log parametrizations. As a result, a
large value of $\kappa$ leads to a large coefficient $h$ 
(\ref{eq:sigma_thinwall_straight_line}) for the Poly-II
parametrization and, consequently, to the behavior of the surface
tension seen in the inset of Fig.~\ref{fig:surface_tension_PolyII}. In the
following, we only discuss the surface tension derived from the Poly-I
and Log parametrizations.

Our results on the zero-temperature surface tension given in Table 
\ref{tab:zeroT_surfacetension} and the temperature dependence of the 
surface tension seen in Fig.~\ref{fig:surface_tension_PolyII} are very 
similar to those found in
Refs.~\cite{Palhares_10,Pinto_12}, which considered the two-flavor QM
and NJL models, respectively. This means that the addition of the
strange sector does not change appreciably the dynamics of the phase
transition at low temperatures and high chemical potentials. Notice
that, given a point of phase coexistence in the $T-\mu$ plane,
$\Delta\sigma_x$ is much larger than $\Delta\sigma_y$, except for
conditions very close to the CEP. Therefore, the coefficient $h$ in
Eq.~(\ref{eq:sigma_thinwall_straight_line}) at low temperatures is
barely changed with the contribution from the strange
sector. Actually, at low temperatures, the Polyakov loop sector does
not give any sensible contribution to the surface tension either.
The temperature range in Fig.~\ref{fig:surface_tension_PolyII} where 
the profiles of the surface tension with the Log and Poly-I parametrizations 
differ indicates where Polyakov loop sector contributes to the surface 
tension and the difference of the profiles can be considered as an estimate 
of the uncertainty of the Polyakov loop contribution.

\begin{table}
  \caption{Low-temperature values of the surface tension at $T=2\,\mrm{MeV}$ 
    (in practice, similar to $T=0$). We choose a sigma meson mass
    $m_\sigma=500\,$MeV and pure glue transition temperature
    $T_0=182\,$MeV. The parameters not shown in this table are found in Table
    \ref{tab:chiral_pot_constants}.}
  \begin{ruledtabular}
  \begin{tabular}{l|c|c|c}
  	Parametrization             			& Log \cite{Roessner_07}		& Poly-I \cite{Scavenius_02}	& Poly-II \cite{Ratti_06}  \\\hline
	$\Sigma_0$ $\mrm{[MeV/fm^2]}$	& $13.0$ 					&  $13.0$ 					& 28.2 \\
  \end{tabular} 
  \end{ruledtabular}
  \label{tab:zeroT_surfacetension}
\end{table}

\subsection{Implications for proto-neutron stars, the early Universe and heavy ion collisions}

The values of surface tension for the $N_f=2+1$ PQM model we found
have interesting implications for several physical scenarios.
For example, compact stars can be considered as laboratories for nuclear 
matter at low temperatures and at such high densities that they may 
contain quark matter \cite{Weber_05}.
Possible scenarios for the formation of quark matter in compact stars are 
old accreting neutron stars, proto-neutron stars after a supernova explosion 
or during the early postbounce evolution of core collapse supernovae \cite{Sagert_09}.
Physical conditions and time scales in these cases imply equilibrium with 
respect to weak interactions and low electron fractions. 
Estimates from \cite{Mintz_10} show that a hadron-quark
phase transition during the bounce phase of a core-collapse supernova
can be dynamically suppressed if the surface tension of this phase
interface is much larger than, say, $20~$MeV/fm$^2$. The estimates
from \cite{Palhares_10,Pinto_12} and this work consistently point
towards low values of the surface tension, which would be compatible
with the formation of quark matter during the bounce.
An observable signal would be a second peak in the neutrino signal 
dominated by the emission of antineutrinos and with a significant change 
in the energy of emitted neutrinos \cite{Sagert_09}. 
However, none
of these calculations really takes into account realistic equations of
state for supernova matter, which has to include not only scalar
mesons, but also vector mesons, nucleons and, very importantly,
leptons. A calculation of surface tension in such a complete model
would be most welcome.

In the cosmological case, physical boundary conditions to describe the QCD 
phase transition in the early Universe include charge neutrality, equilibrium 
with respect to weak interactions and baryon and lepton asymmetries consistent 
with observations. Observations of the cosmic microwave background radiation 
and constraints from primordial nucleosynthesis require a tiny baryon asymmetry 
at relatively low temperatures ($T\lesssim 1\,\mrm{MeV}$) \cite{Boyanovsky_06}. 
In the standard scenario this observational constraint is extrapolated up to the 
scale of the QCD phase transition that is then a smooth crossover. 
In the scenario of little inflation~\cite{Boeckel_10},
the Universe enters the QCD era with a very high quark density. As a
result, the quark-gluon plasma (QGP) that fills the Universe cools
down until it eventually crosses a first-order line of the phase
diagram, becoming metastable. While the QGP is metastable, the
conditions for a cosmic inflation can be met and the extra quark
density becomes very diluted.
Observational signals of this evolution include an enhancement of primordial 
density fluctuations on stellar up to galactic scales, production of galactic 
and extragalactic magnetic fields and a modification of the gravitational wave 
spectrum~\cite{Boeckel_12,Schettler_11}. 
At some point of the
expansion, however, the phase transition from a QGP to hadronic
matter must happen, most likely through bubble nucleation. In order to
be effective, the baryon dilution required by the little inflation
scenario needs to be long enough. This requires that the QGP remains
metastable even for high degrees of supercooling, something that
requires a very low nucleation rate and, therefore, a large value of
surface tension. As we have discussed, however, the values of surface
tension found with chiral models, including the study we have performed in this
work, are relatively small and possibly do
not allow the strong metastability required by the scenario of little
inflation.
However, also a large lepton asymmetry can drive the evolutionary path of 
the Universe towards larger quark chemical potentials~\cite{Schwarz_09}.

In heavy ion collisions the phase boundary of hadronic and quark matter is, 
if at all, crossed twice. First, as formation of a quark gluon plasma in a 
hadronic gas and then as rehadronization of the fireball. Complications in 
studying nucleation in heavy ion collisions are the short time scales and 
the finite size of the system. Additionally, the nucleation rate has to be 
considered in relation to the expansion time. These conditions can lead to 
the fact that the system stays in the metastable state close to the spinodal 
instability and that the dominant mechanism for phase conversion is the 
alternative scenario to homogenous nucleation, namely spinodal decomposition 
\cite{Scavenius_01}.
Nevertheless, the growth rate of fluctuations by spinodal instabilities 
is closely related to the surface tension
\cite{Randrup_09, Steinheimer_12}. They can lead to observable signatures 
for the value of the surface tension we found \cite{Steinheimer_12}.
Additionally, these fluctuations can be amplified by nucleation in the metastable region.
The relatively small values of the surface tension we found suggest a early 
nucleation of small quark-gluon plasma droplets at relatively modest energies 
like at FAIR's SIS 100 \cite{Kapusta_95}.
The details of rehadronization leave their fingerprints on those observables 
that are sensitive to the life-time of the fireball. Weak supercooling favors 
the thermal freeze-out to happen in the hadronic phase with impact on particle 
yields and spectra \cite{Csernai_95} and a distinct hydrodynamic expansion 
pattern \cite{Scavenius_01}.

\section{\label{sec5}Conclusion} 

In this work, we have analyzed some thermodynamical properties of the
PQM model with $N_f=2+1$ constituent quarks. In particular, we were
concerned about the definition of the equilibrium state of a system at
temperature $T$ and quark chemical potential $\mu$. As in the case of
the PNJL model, the in-medium effective potential of the PQM model is
not a real function of real variables and, therefore, it has no
minima. As a way to circumvent this problem, we have first rewritten
the effective potential in terms of real variables only, such that
the real and imaginary parts of the potential are separated. The
relation between the imaginary part of the effective potential and
the fermion sign problem is discussed. The real effective potential is
found to be the appropriate quantity to be minimized and it is a
consistent quantity in terms of standard arguments of equilibrium
statistical physics. This allowed us to calculate some properties of
the model in equilibrium at finite $T$ and $\mu$. In particular, we
calculated the evolution of the order parameters as functions of the
temperature (for $\mu=0$) and of the quark chemical potential (for
$T=2\,\mrm{MeV}$, which is, in practice, equivalent to $T=0$). We also
calculated the $T-\mu$ phase diagram of the model and compared our
results to the ones found in the literature using the saddle-point
method. The results are equivalent for $\mu=0$ (where they had to
be), but differ for nonzero $\mu$. 
The phase transition lines of both methods coincide but the location of 
the critical end point can be at up to 40\% larger chemical potentials 
using the minimizing method depending on the Polyakov loop potential 
parametrization and parameters.

A careful minimization of the effective potential also allowed us to
study the problem of homogeneous nucleation of bubbles in a
first-order phase transition. More specifically, we calculated the
surface tension of an interface between the two phases predicted by
the model, a quantity that is crucial for the nucleation rate of
bubbles in a first-order phase transition. We saw that the $N_f=2+1$
PQM model yields results very similar to those of the two-flavor NJL
and QM models, so that the influence of both the strange quark and
the Polyakov loop at low temperatures is small. However, the same
cannot be said for higher temperatures, where all contributions become
of the same order of magnitude. Our overestimate gives a
conservative upper bound of $\Sigma\lesssim15~$MeV/fm$^2$ for the
surface tension. The actual value, however, may be even lower, not
only because of the direct approximations in the calculation of the
surface tension, but also due to vacuum terms, which we have
neglected and should make the first-order transition weaker at low $T$
and high $\mu$. In any case, this reinforces the trend shown
recently in Refs.~\cite{Palhares_10,Pinto_12} of a low surface tension
in chiral models for QCD at finite baryon density. Such a low value
would allow a quick hadron-quark phase conversion.
This implies interesting implications for several physical scenarios,
be it heavy ion collisions, proto-neutron stars or the early Universe.

In summary, the analysis carried out in this work suggests that, in
spite of the good agreement of the chiral models at finite $T$ and
$\mu=0$ with lattice calculations, one should take care when the
$\mu\not=0$ case is addressed. This difficulty is manifest in the
non-reality of the equilibrium effective potential for $\mu\not=0$,
even far from any phase transition. Notice that, in
principle, the same problem can affect any chiral model with gauge
fields coupled to quarks at finite $\mu$, such as the PNJL model. We
believe that more consistent solutions to these (and other possible)
problems are still needed as they would bring more confidence to
further progress in the domain of high chemical potentials of the
QCD phase diagram with effective models.

\begin{acknowledgments}

We thank Marcus B. Pinto, Eduardo S.~Fraga, Let\'\i cia
F.~Palhares and Jan M.~Pawlowski for discussions. The work of B.W.M.\ and R.O.R.\ was
partially supported by Conselho Nacional de Desenvolvimento
Cient\'{\i}fico e Tecnol\'ogico (CNPq). R.O.R.\ is also partially
supported by a research grant from Funda\c{c}\~ao Carlos Chagas Filho
de Amparo \`a Pesquisa do Estado do Rio de Janeiro (FAPERJ). 
J.S.B.\ and R.S.\ are supported by BMBF under grants FKZ 06HD9127 and 06HD7142, by the German Research Foundation (DFG) within the framework of the excellence initiative through the Heidelberg Graduate School of Fundamental Physics (HGSFP) and through the Helmholtz Graduate School for Heavy-Ion Research (HGS-HIRe) and the Graduate Program for Hadron and Ion Research (GP-HIR) by the Gesellschaft f\"ur Schwer\-ionen\-forschung (GSI), Darmstadt and the Alliance Program of the Helmholtz Association HA216/EMMI.
R.S.\ thanks for the kind hospitality and support at the Instituto de F\'\i sica at the Universidade Federal do Rio de Janeiro during part of this project, a research stay that was also supported by the HGS-HIRe abroad program.

\end{acknowledgments}

\bibliography{references} 

\begin{thebibliography}{10}%
\makeatletter
\providecommand \@ifxundefined [1]{%
 \ifx #1\undefined \expandafter \@firstoftwo
 \else \expandafter \@secondoftwo
\fi
}%
\providecommand \@ifnum [1]{%
 \ifnum #1\expandafter \@firstoftwo
 \else \expandafter \@secondoftwo
\fi
}%
\providecommand \enquote [1]{``#1''}%
\providecommand \bibnamefont  [1]{#1}%
\providecommand \bibfnamefont [1]{#1}%
\providecommand \citenamefont [1]{#1}%
\providecommand\href[0]{\@sanitize\@href}%
\providecommand\@href[1]{\endgroup\@@startlink{#1}\endgroup\@@href}%
\providecommand\@@href[1]{#1\@@endlink}%
\providecommand \@sanitize [0]{\begingroup\catcode`\&12\catcode`\#12\relax}%
\@ifxundefined \pdfoutput {\@firstoftwo}{%
 \@ifnum{\z@=\pdfoutput}{\@firstoftwo}{\@secondoftwo}%
}{%
 \providecommand\@@startlink[1]{\leavevmode\special{html:<a href="#1">}}%
 \providecommand\@@endlink[0]{\special{html:</a>}}%
}{%
 \providecommand\@@startlink[1]{%
  \leavevmode
  \pdfstartlink
   attr{/Border[0 0 1 ]/H/I/C[0 1 1]}%
   user{/Subtype/Link/A<</Type/Action/S/URI/URI(#1)>>}%
  \relax
 }%
 \providecommand\@@endlink[0]{\pdfendlink}%
}%
\providecommand \url  [0]{\begingroup\@sanitize \@url }%
\providecommand \@url [1]{\endgroup\@href {#1}{\urlprefix}}%
\providecommand \urlprefix [0]{URL }%
\providecommand \Eprint[0]{\href }%
\@ifxundefined \urlstyle {%
  \providecommand \doi [1]{doi:\discretionary{}{}{}#1}%
}{%
  \providecommand \doi [0]{doi:\discretionary{}{}{}\begingroup
  \urlstyle{rm}\Url }%
}%
\providecommand \doibase [0]{http://dx.doi.org/}%
\providecommand \Doi[1]{\href{\doibase#1}}%
\providecommand \bibAnnote [3]{%
  \BibitemShut{#1}%
  \begin{quotation}\noindent
    \textsc{Key:}\ #2\\\textsc{Annotation:}\ #3%
  \end{quotation}%
}%
\providecommand \bibAnnoteFile [2]{%
  \IfFileExists{#2}{\bibAnnote {#1} {#2} {\input{#2}}}{}%
}%
\providecommand \typeout [0]{\immediate \write \m@ne }%
\providecommand \selectlanguage [0]{\@gobble}%
\providecommand \bibinfo [0]{\@secondoftwo}%
\providecommand \bibfield [0]{\@secondoftwo}%
\providecommand \translation [1]{[#1]}%
\providecommand \BibitemOpen[0]{}%
\providecommand \bibitemStop [0]{}%
\providecommand \bibitemNoStop [0]{.\EOS\space}%
\providecommand \EOS [0]{\spacefactor3000\relax}%
\providecommand \BibitemShut [1]{\csname bibitem#1\endcsname}%
\bibitem{deForcrand_06}%
  \BibitemOpen
  \bibfield{author}{%
  \bibinfo {author} {\bibfnamefont{P.}~\bibnamefont{{de Forcrand}}}\ and\
  \bibinfo {author} {\bibfnamefont{O.}~\bibnamefont{{Philipsen}}},\ }%
  \bibfield{journal}{%
  \Doi{10.1088/1126-6708/2007/01/077}{\bibinfo {journal} {Journal of High
  Energy Physics}}\ }%
  \textbf{\bibinfo {volume} {1}},\ \bibinfo {eid} {077} (\bibinfo {year}
  {2007}),\
  \Eprint{http://arxiv.org/abs/arXiv:hep-lat/0607017}{arXiv:hep-lat/0607017}%
  \bibAnnoteFile{NoStop}{deForcrand_06}%
\bibitem{DeTar_09}%
  \BibitemOpen
  \bibfield{author}{%
  \bibinfo {author} {\bibfnamefont{C.~E.}\ \bibnamefont{{Detar}}}\ and\
  \bibinfo {author} {\bibfnamefont{U.~M.}\ \bibnamefont{{Heller}}},\ }%
  \bibfield{journal}{%
  \Doi{10.1140/epja/i2009-10825-3}{\bibinfo {journal} {Eur. Phys. Journ. A}}\
  }%
  \textbf{\bibinfo {volume} {41}},\ \bibinfo {pages} {405} (\bibinfo {year}
  {2009}),\ \Eprint{http://arxiv.org/abs/0905.2949}{arXiv:0905.2949 [hep-lat]}%
  \bibAnnoteFile{NoStop}{DeTar_09}%
\bibitem{Endrodi:2011gv}%
  \BibitemOpen
  \bibfield{author}{%
  \bibinfo {author} {\bibfnamefont{G.}~\bibnamefont{Endrodi}}, \bibinfo
  {author} {\bibfnamefont{Z.}~\bibnamefont{Fodor}}, \bibinfo {author}
  {\bibfnamefont{S.}~\bibnamefont{Katz}},\ and\ \bibinfo {author}
  {\bibfnamefont{K.}~\bibnamefont{Szabo}},\ }%
  \bibfield{journal}{%
  \Doi{10.1007/JHEP04(2011)001}{\bibinfo {journal} {JHEP}}\ }%
  \textbf{\bibinfo {volume} {1104}},\ \bibinfo {pages} {001} (\bibinfo {year}
  {2011}),\ \Eprint{http://arxiv.org/abs/1102.1356}{arXiv:1102.1356 [hep-lat]}%
  \bibAnnoteFile{NoStop}{Endrodi:2011gv}%
\bibitem{Nambu_61_A}%
  \BibitemOpen
  \bibfield{author}{%
  \bibinfo {author} {\bibfnamefont{Y.}~\bibnamefont{Nambu}}\ and\ \bibinfo
  {author} {\bibfnamefont{G.}~\bibnamefont{Jona-Lasinio}},\ }%
  \bibfield{journal}{%
  \Doi{10.1103/PhysRev.122.345}{\bibinfo {journal} {Phys. Rev.}}\ }%
  \textbf{\bibinfo {volume} {122}},\ \bibinfo {pages} {345} (\bibinfo {year}
  {1961})%
  \bibAnnoteFile{NoStop}{Nambu_61_A}%
\bibitem{GellMann_60}%
  \BibitemOpen
  \bibfield{author}{%
  \bibinfo {author} {\bibfnamefont{M.}~\bibnamefont{Gell-Mann}}\ and\ \bibinfo
  {author} {\bibfnamefont{M.}~\bibnamefont{Levy}},\ }%
  \bibfield{journal}{%
  \Doi{10.1007/BF02859738}{\bibinfo {journal} {Nuovo Cim.}}\ }%
  \textbf{\bibinfo {volume} {16}},\ \bibinfo {pages} {705} (\bibinfo {year}
  {1960})%
  \bibAnnoteFile{NoStop}{GellMann_60}%
\bibitem{Schaefer_07}%
  \BibitemOpen
  \bibfield{author}{%
  \bibinfo {author} {\bibfnamefont{B.}~\bibnamefont{{Schaefer}}}, \bibinfo
  {author} {\bibfnamefont{J.~M.}\ \bibnamefont{{Pawlowski}}},\ and\ \bibinfo
  {author} {\bibfnamefont{J.}~\bibnamefont{{Wambach}}},\ }%
  \bibfield{journal}{%
  \Doi{10.1103/PhysRevD.76.074023}{\bibinfo {journal} {\prd}}\ }%
  \textbf{\bibinfo {volume} {76}},\ \bibinfo {pages} {074023} (\bibinfo {year}
  {2007}),\ \Eprint{http://arxiv.org/abs/0704.3234}{arXiv:0704.3234 [hep-ph]}%
  \bibAnnoteFile{NoStop}{Schaefer_07}%
\bibitem{Schaefer_10}%
  \BibitemOpen
  \bibfield{author}{%
  \bibinfo {author} {\bibfnamefont{B.}~\bibnamefont{{Schaefer}}}, \bibinfo
  {author} {\bibfnamefont{M.}~\bibnamefont{{Wagner}}},\ and\ \bibinfo {author}
  {\bibfnamefont{J.}~\bibnamefont{{Wambach}}},\ }%
  \bibfield{journal}{%
  \Doi{10.1103/PhysRevD.81.074013}{\bibinfo {journal} {\prd}}\ }%
  \textbf{\bibinfo {volume} {81}},\ \bibinfo {pages} {074013} (\bibinfo {year}
  {2010}),\ \Eprint{http://arxiv.org/abs/0910.5628}{arXiv:0910.5628 [hep-ph]}%
  \bibAnnoteFile{NoStop}{Schaefer_10}%
\bibitem{Herbst_11}%
  \BibitemOpen
  \bibfield{author}{%
  \bibinfo {author} {\bibfnamefont{T.~K.}\ \bibnamefont{{Herbst}}}, \bibinfo
  {author} {\bibfnamefont{J.~M.}\ \bibnamefont{{Pawlowski}}},\ and\ \bibinfo
  {author} {\bibfnamefont{B.}~\bibnamefont{{Schaefer}}},\ }%
  \bibfield{journal}{%
  \Doi{10.1016/j.physletb.2010.12.003}{\bibinfo {journal} {Phys. Lett. B}}\ }%
  \textbf{\bibinfo {volume} {696}},\ \bibinfo {pages} {58} (\bibinfo {year}
  {2011}),\ \Eprint{http://arxiv.org/abs/1008.0081}{arXiv:1008.0081 [hep-ph]}%
  \bibAnnoteFile{NoStop}{Herbst_11}%
\bibitem{Schaefer_12}%
  \BibitemOpen
  \bibfield{author}{%
  \bibinfo {author} {\bibfnamefont{B.-J.}\ \bibnamefont{{Schaefer}}}\ and\
  \bibinfo {author} {\bibfnamefont{M.}~\bibnamefont{{Wagner}}},\ }%
  \bibfield{journal}{%
  \Doi{10.1103/PhysRevD.85.034027}{\bibinfo {journal} {\prd}}\ }%
  \textbf{\bibinfo {volume} {85}},\ \bibinfo {eid} {034027} (\bibinfo {year}
  {2012}),\ \Eprint{http://arxiv.org/abs/1111.6871}{arXiv:1111.6871 [hep-ph]}%
  \bibAnnoteFile{NoStop}{Schaefer_12}%
\bibitem{Haas:2013qwp}%
  \BibitemOpen
  \bibfield{author}{%
  \bibinfo {author} {\bibfnamefont{L.~M.}\ \bibnamefont{Haas}}, \bibinfo
  {author} {\bibfnamefont{R.}~\bibnamefont{Stiele}}, \bibinfo {author}
  {\bibfnamefont{J.}~\bibnamefont{Braun}}, \bibinfo {author}
  {\bibfnamefont{J.~M.}\ \bibnamefont{Pawlowski}},\ and\ \bibinfo {author}
  {\bibfnamefont{J.}~\bibnamefont{Schaffner-Bielich}}}%
   (\bibinfo {year} {2013}),\
  \Eprint{http://arxiv.org/abs/1302.1993}{arXiv:1302.1993 [hep-ph]}%
  \bibAnnoteFile{NoStop}{Haas:2013qwp}%
\bibitem{Stiele:2013}%
  \BibitemOpen
  \bibfield{author}{%
  \bibinfo {author} {\bibfnamefont{R.}~\bibnamefont{Stiele}}, \bibinfo {author}
  {\bibfnamefont{L.~M.}\ \bibnamefont{Haas}}, \bibinfo {author}
  {\bibfnamefont{J.}~\bibnamefont{Braun}}, \bibinfo {author}
  {\bibfnamefont{J.~M.}\ \bibnamefont{Pawlowski}},\ and\ \bibinfo {author}
  {\bibfnamefont{J.}~\bibnamefont{Schaffner-Bielich}},\ }%
  \enquote{\bibinfo {title} {{QCD thermodynamics of effective models with an
  improved Polyakov-loop potential}},}\  (\bibinfo {year} {2013}),\ \bibinfo
  {note} {contribution to the POS Proceedings of the Xth Quark Confinement and
  the Hadron Spectrum conference, PoS(Confinement X)215}%
  \bibAnnoteFile{NoStop}{Stiele:2013}%
\bibitem{Aoki:2006we}%
  \BibitemOpen
  \bibfield{author}{%
  \bibinfo {author} {\bibfnamefont{Y.}~\bibnamefont{Aoki}}, \bibinfo {author}
  {\bibfnamefont{G.}~\bibnamefont{Endrodi}}, \bibinfo {author}
  {\bibfnamefont{Z.}~\bibnamefont{Fodor}}, \bibinfo {author}
  {\bibfnamefont{S.}~\bibnamefont{Katz}},\ and\ \bibinfo {author}
  {\bibfnamefont{K.}~\bibnamefont{Szabo}},\ }%
  \bibfield{journal}{%
  \Doi{10.1038/nature05120}{\bibinfo {journal} {Nature}}\ }%
  \textbf{\bibinfo {volume} {443}},\ \bibinfo {pages} {675} (\bibinfo {year}
  {2006}),\ \Eprint{http://arxiv.org/abs/hep-lat/0611014}{arXiv:hep-lat/0611014
  [hep-lat]}%
  \bibAnnoteFile{NoStop}{Aoki:2006we}%
\bibitem{Fukushima_07}%
  \BibitemOpen
  \bibfield{author}{%
  \bibinfo {author} {\bibfnamefont{K.}~\bibnamefont{{Fukushima}}}\ and\
  \bibinfo {author} {\bibfnamefont{Y.}~\bibnamefont{{Hidaka}}},\ }%
  \bibfield{journal}{%
  \Doi{10.1103/PhysRevD.75.036002}{\bibinfo {journal} {\prd}}\ }%
  \textbf{\bibinfo {volume} {75}},\ \bibinfo {eid} {036002} (\bibinfo {year}
  {2007}),\
  \Eprint{http://arxiv.org/abs/arXiv:hep-ph/0610323}{arXiv:hep-ph/0610323}%
  \bibAnnoteFile{NoStop}{Fukushima_07}%
\bibitem{Dumitru_05}%
  \BibitemOpen
  \bibfield{author}{%
  \bibinfo {author} {\bibfnamefont{A.}~\bibnamefont{{Dumitru}}}, \bibinfo
  {author} {\bibfnamefont{R.~D.}\ \bibnamefont{{Pisarski}}},\ and\ \bibinfo
  {author} {\bibfnamefont{D.}~\bibnamefont{{Zschiesche}}},\ }%
  \bibfield{journal}{%
  \Doi{10.1103/PhysRevD.72.065008}{\bibinfo {journal} {\prd}}\ }%
  \textbf{\bibinfo {volume} {72}},\ \bibinfo {eid} {065008} (\bibinfo {year}
  {2005}),\
  \Eprint{http://arxiv.org/abs/arXiv:hep-ph/0505256}{arXiv:hep-ph/0505256}%
  \bibAnnoteFile{NoStop}{Dumitru_05}%
\bibitem{Roessner_08}%
  \BibitemOpen
  \bibfield{author}{%
  \bibinfo {author} {\bibfnamefont{S.}~\bibnamefont{{R{\"o}{\ss}ner}}},
  \bibinfo {author} {\bibfnamefont{T.}~\bibnamefont{{Hell}}}, \bibinfo {author}
  {\bibfnamefont{C.}~\bibnamefont{{Ratti}}},\ and\ \bibinfo {author}
  {\bibfnamefont{W.}~\bibnamefont{{Weise}}},\ }%
  \bibfield{journal}{%
  \Doi{10.1016/j.nuclphysa.2008.10.006}{\bibinfo {journal} {Nucl. Phys. A}}\ }%
  \textbf{\bibinfo {volume} {814}},\ \bibinfo {pages} {118} (\bibinfo {year}
  {2008}),\ \Eprint{http://arxiv.org/abs/0712.3152}{arXiv:0712.3152 [hep-ph]}%
  \bibAnnoteFile{NoStop}{Roessner_08}%
\bibitem{Fukushima_04}%
  \BibitemOpen
  \bibfield{author}{%
  \bibinfo {author} {\bibfnamefont{K.}~\bibnamefont{{Fukushima}}},\ }%
  \bibfield{journal}{%
  \Doi{10.1016/j.physletb.2004.04.027}{\bibinfo {journal} {Phys. Lett. B}}\ }%
  \textbf{\bibinfo {volume} {591}},\ \bibinfo {pages} {277} (\bibinfo {year}
  {2004}),\
  \Eprint{http://arxiv.org/abs/arXiv:hep-ph/0310121}{arXiv:hep-ph/0310121}%
  \bibAnnoteFile{NoStop}{Fukushima_04}%
\bibitem{Scavenius_02}%
  \BibitemOpen
  \bibfield{author}{%
  \bibinfo {author} {\bibfnamefont{O.}~\bibnamefont{{Scavenius}}}, \bibinfo
  {author} {\bibfnamefont{A.}~\bibnamefont{{Dumitru}}},\ and\ \bibinfo {author}
  {\bibfnamefont{J.~T.}\ \bibnamefont{{Lenaghan}}},\ }%
  \bibfield{journal}{%
  \Doi{10.1103/PhysRevC.66.034903}{\bibinfo {journal} {\prc}}\ }%
  \textbf{\bibinfo {volume} {66}},\ \bibinfo {eid} {034903} (\bibinfo {year}
  {2002}),\
  \Eprint{http://arxiv.org/abs/arXiv:hep-ph/0201079}{arXiv:hep-ph/0201079}%
  \bibAnnoteFile{NoStop}{Scavenius_02}%
\bibitem{Ratti_06}%
  \BibitemOpen
  \bibfield{author}{%
  \bibinfo {author} {\bibfnamefont{C.}~\bibnamefont{{Ratti}}}, \bibinfo
  {author} {\bibfnamefont{M.~A.}\ \bibnamefont{{Thaler}}},\ and\ \bibinfo
  {author} {\bibfnamefont{W.}~\bibnamefont{{Weise}}},\ }%
  \bibfield{journal}{%
  \Doi{10.1103/PhysRevD.73.014019}{\bibinfo {journal} {\prd}}\ }%
  \textbf{\bibinfo {volume} {73}},\ \bibinfo {pages} {014019} (\bibinfo {year}
  {2006}),\
  \Eprint{http://arxiv.org/abs/arXiv:hep-ph/0506234}{arXiv:hep-ph/0506234}%
  \bibAnnoteFile{NoStop}{Ratti_06}%
\bibitem{Roessner_07}%
  \BibitemOpen
  \bibfield{author}{%
  \bibinfo {author} {\bibfnamefont{S.}~\bibnamefont{{R{\"o}{\ss}ner}}},
  \bibinfo {author} {\bibfnamefont{C.}~\bibnamefont{{Ratti}}},\ and\ \bibinfo
  {author} {\bibfnamefont{W.}~\bibnamefont{{Weise}}},\ }%
  \bibfield{journal}{%
  \Doi{10.1103/PhysRevD.75.034007}{\bibinfo {journal} {\prd}}\ }%
  \textbf{\bibinfo {volume} {75}},\ \bibinfo {eid} {034007} (\bibinfo {year}
  {2007}),\
  \Eprint{http://arxiv.org/abs/arXiv:hep-ph/0609281}{arXiv:hep-ph/0609281}%
  \bibAnnoteFile{NoStop}{Roessner_07}%
\bibitem{Langer_73}%
  \BibitemOpen
  \bibfield{author}{%
  \bibinfo {author} {\bibfnamefont{J.~S.}\ \bibnamefont{{Langer}}}\ and\
  \bibinfo {author} {\bibfnamefont{L.~A.}\ \bibnamefont{{Turski}}},\ }%
  \bibfield{journal}{%
  \Doi{10.1103/PhysRevA.8.3230}{\bibinfo {journal} {\pra}}\ }%
  \textbf{\bibinfo {volume} {8}},\ \bibinfo {pages} {3230} (\bibinfo {year}
  {1973})%
  \bibAnnoteFile{NoStop}{Langer_73}%
\bibitem{Csernai:1992tj}%
  \BibitemOpen
  \bibfield{author}{%
  \bibinfo {author} {\bibfnamefont{L.~P.}\ \bibnamefont{Csernai}}\ and\
  \bibinfo {author} {\bibfnamefont{J.~I.}\ \bibnamefont{Kapusta}},\ }%
  \bibfield{journal}{%
  \Doi{10.1103/PhysRevD.46.1379}{\bibinfo {journal} {Phys. Rev.}}\ }%
  \textbf{\bibinfo {volume} {D46}},\ \bibinfo {pages} {1379} (\bibinfo {year}
  {1992})%
  \bibAnnoteFile{NoStop}{Csernai:1992tj}%
\bibitem{Gleiser:1993hf}%
  \BibitemOpen
  \bibfield{author}{%
  \bibinfo {author} {\bibfnamefont{M.}~\bibnamefont{Gleiser}}, \bibinfo
  {author} {\bibfnamefont{G.~C.}\ \bibnamefont{Marques}},\ and\ \bibinfo
  {author} {\bibfnamefont{R.~O.}\ \bibnamefont{Ramos}},\ }%
  \bibfield{journal}{%
  \Doi{10.1103/PhysRevD.48.1571}{\bibinfo {journal} {Phys. Rev.}}\ }%
  \textbf{\bibinfo {volume} {D48}},\ \bibinfo {pages} {1571} (\bibinfo {year}
  {1993}),\ \Eprint{http://arxiv.org/abs/hep-ph/9304234}{arXiv:hep-ph/9304234
  [hep-ph]}%
  \bibAnnoteFile{NoStop}{Gleiser:1993hf}%
\bibitem{Scavenius_01}%
  \BibitemOpen
  \bibfield{author}{%
  \bibinfo {author} {\bibfnamefont{O.}~\bibnamefont{{Scavenius}}}, \bibinfo
  {author} {\bibfnamefont{A.}~\bibnamefont{{Dumitru}}}, \bibinfo {author}
  {\bibfnamefont{E.~S.}\ \bibnamefont{{Fraga}}}, \bibinfo {author}
  {\bibfnamefont{J.~T.}\ \bibnamefont{{Lenaghan}}},\ and\ \bibinfo {author}
  {\bibfnamefont{A.~D.}\ \bibnamefont{{Jackson}}},\ }%
  \bibfield{journal}{%
  \Doi{10.1103/PhysRevD.63.116003}{\bibinfo {journal} {\prd}}\ }%
  \textbf{\bibinfo {volume} {63}},\ \bibinfo {eid} {116003} (\bibinfo {year}
  {2001}),\
  \Eprint{http://arxiv.org/abs/arXiv:hep-ph/0009171}{arXiv:hep-ph/0009171}%
  \bibAnnoteFile{NoStop}{Scavenius_01}%
\bibitem{Bessa_09}%
  \BibitemOpen
  \bibfield{author}{%
  \bibinfo {author} {\bibfnamefont{A.}~\bibnamefont{{Bessa}}}, \bibinfo
  {author} {\bibfnamefont{E.~S.}\ \bibnamefont{{Fraga}}},\ and\ \bibinfo
  {author} {\bibfnamefont{B.~W.}\ \bibnamefont{{Mintz}}},\ }%
  \bibfield{journal}{%
  \Doi{10.1103/PhysRevD.79.034012}{\bibinfo {journal} {\prd}}\ }%
  \textbf{\bibinfo {volume} {79}},\ \bibinfo {eid} {034012} (\bibinfo {year}
  {2009}),\ \Eprint{http://arxiv.org/abs/0811.4385}{arXiv:0811.4385 [hep-ph]}%
  \bibAnnoteFile{NoStop}{Bessa_09}%
\bibitem{Bombaci:2008wg}%
  \BibitemOpen
  \bibfield{author}{%
  \bibinfo {author} {\bibfnamefont{I.}~\bibnamefont{{Bombaci}}}, \bibinfo
  {author} {\bibfnamefont{P.~K.}\ \bibnamefont{{Panda}}}, \bibinfo {author}
  {\bibfnamefont{C.}~\bibnamefont{{Provid{\^e}ncia}}},\ and\ \bibinfo {author}
  {\bibfnamefont{I.}~\bibnamefont{{Vida{\~n}a}}},\ }%
  \bibfield{journal}{%
  \Doi{10.1103/PhysRevD.77.083002}{\bibinfo {journal} {\prd}}\ }%
  \textbf{\bibinfo {volume} {77}},\ \bibinfo {eid} {083002} (\bibinfo {year}
  {2008}),\ \Eprint{http://arxiv.org/abs/0802.1794}{arXiv:0802.1794
  [astro-ph]}%
  \bibAnnoteFile{NoStop}{Bombaci:2008wg}%
\bibitem{Bombaci:2009jt}%
  \BibitemOpen
  \bibfield{author}{%
  \bibinfo {author} {\bibfnamefont{I.}~\bibnamefont{{Bombaci}}}, \bibinfo
  {author} {\bibfnamefont{D.}~\bibnamefont{{Logoteta}}}, \bibinfo {author}
  {\bibfnamefont{P.~K.}\ \bibnamefont{{Panda}}}, \bibinfo {author}
  {\bibfnamefont{C.}~\bibnamefont{{Provid{\^e}ncia}}},\ and\ \bibinfo {author}
  {\bibfnamefont{I.}~\bibnamefont{{Vida{\~n}a}}},\ }%
  \bibfield{journal}{%
  \Doi{10.1016/j.physletb.2009.09.039}{\bibinfo {journal} {Physics Letters B}}\
  }%
  \textbf{\bibinfo {volume} {680}},\ \bibinfo {pages} {448} (\bibinfo {year}
  {2009}),\ \Eprint{http://arxiv.org/abs/0910.4109}{arXiv:0910.4109
  [astro-ph.SR]}%
  \bibAnnoteFile{NoStop}{Bombaci:2009jt}%
\bibitem{Mintz_10}%
  \BibitemOpen
  \bibfield{author}{%
  \bibinfo {author} {\bibfnamefont{B.~W.}\ \bibnamefont{{Mintz}}}, \bibinfo
  {author} {\bibfnamefont{E.~S.}\ \bibnamefont{{Fraga}}}, \bibinfo {author}
  {\bibfnamefont{G.}~\bibnamefont{{Pagliara}}},\ and\ \bibinfo {author}
  {\bibfnamefont{J.}~\bibnamefont{{Schaffner-Bielich}}},\ }%
  \bibfield{journal}{%
  \Doi{10.1103/PhysRevD.81.123012}{\bibinfo {journal} {\prd}}\ }%
  \textbf{\bibinfo {volume} {81}},\ \bibinfo {eid} {123012} (\bibinfo {year}
  {2010}),\ \Eprint{http://arxiv.org/abs/0910.3927}{arXiv:0910.3927 [hep-ph]}%
  \bibAnnoteFile{NoStop}{Mintz_10}%
\bibitem{Boeckel_10}%
  \BibitemOpen
  \bibfield{author}{%
  \bibinfo {author} {\bibfnamefont{T.}~\bibnamefont{{Boeckel}}}\ and\ \bibinfo
  {author} {\bibfnamefont{J.}~\bibnamefont{{Schaffner-Bielich}}},\ }%
  \bibfield{journal}{%
  \Doi{10.1103/PhysRevLett.105.041301}{\bibinfo {journal} {Phys.~Rev.~Lett.}}\
  }%
  \textbf{\bibinfo {volume} {105}},\ \bibinfo {eid} {041301} (\bibinfo {year}
  {2010}),\ \Eprint{http://arxiv.org/abs/0906.4520}{arXiv:0906.4520
  [astro-ph.CO]}%
  \bibAnnoteFile{NoStop}{Boeckel_10}%
\bibitem{Boeckel_12}%
  \BibitemOpen
  \bibfield{author}{%
  \bibinfo {author} {\bibfnamefont{T.}~\bibnamefont{{Boeckel}}}\ and\ \bibinfo
  {author} {\bibfnamefont{J.}~\bibnamefont{{Schaffner-Bielich}}},\ }%
  \bibfield{journal}{%
  \Doi{10.1103/PhysRevD.85.103506}{\bibinfo {journal} {\prd}}\ }%
  \textbf{\bibinfo {volume} {85}},\ \bibinfo {eid} {103506} (\bibinfo {year}
  {2012}),\ \Eprint{http://arxiv.org/abs/1105.0832}{arXiv:1105.0832
  [astro-ph.CO]}%
  \bibAnnoteFile{NoStop}{Boeckel_12}%
\bibitem{Palhares_10}%
  \BibitemOpen
  \bibfield{author}{%
  \bibinfo {author} {\bibfnamefont{L.~F.}\ \bibnamefont{{Palhares}}}\ and\
  \bibinfo {author} {\bibfnamefont{E.~S.}\ \bibnamefont{{Fraga}}},\ }%
  \bibfield{journal}{%
  \Doi{10.1103/PhysRevD.82.125018}{\bibinfo {journal} {\prd}}\ }%
  \textbf{\bibinfo {volume} {82}},\ \bibinfo {eid} {125018} (\bibinfo {year}
  {2010}),\ \Eprint{http://arxiv.org/abs/1006.2357}{arXiv:1006.2357 [hep-ph]}%
  \bibAnnoteFile{NoStop}{Palhares_10}%
\bibitem{Pinto_12}%
  \BibitemOpen
  \bibfield{author}{%
  \bibinfo {author} {\bibfnamefont{M.~B.}\ \bibnamefont{{Pinto}}}, \bibinfo
  {author} {\bibfnamefont{V.}~\bibnamefont{{Koch}}},\ and\ \bibinfo {author}
  {\bibfnamefont{J.}~\bibnamefont{{Randrup}}},\ }%
  \bibfield{journal}{%
  \Doi{10.1103/PhysRevC.86.025203}{\bibinfo {journal} {\prc}}\ }%
  \textbf{\bibinfo {volume} {86}},\ \bibinfo {eid} {025203} (\bibinfo {year}
  {2012}),\ \Eprint{http://arxiv.org/abs/1207.5186}{arXiv:1207.5186 [hep-ph]}%
  \bibAnnoteFile{NoStop}{Pinto_12}%
\bibitem{Mocsy_04}%
  \BibitemOpen
  \bibfield{author}{%
  \bibinfo {author} {\bibfnamefont{{\'A}.}~\bibnamefont{{M{\'o}csy}}}, \bibinfo
  {author} {\bibfnamefont{F.}~\bibnamefont{{Sannino}}},\ and\ \bibinfo {author}
  {\bibfnamefont{K.}~\bibnamefont{{Tuominen}}},\ }%
  \bibfield{journal}{%
  \Doi{10.1103/PhysRevLett.92.182302}{\bibinfo {journal} {Phys. Rev. Lett.}}\
  }%
  \textbf{\bibinfo {volume} {92}},\ \bibinfo {pages} {182302} (\bibinfo {year}
  {2004}),\
  \Eprint{http://arxiv.org/abs/arXiv:hep-ph/0308135}{arXiv:hep-ph/0308135}%
  \bibAnnoteFile{NoStop}{Mocsy_04}%
\bibitem{Megias_06_A}%
  \BibitemOpen
  \bibfield{author}{%
  \bibinfo {author} {\bibfnamefont{E.}~\bibnamefont{{Meg{\'{\i}}as}}}, \bibinfo
  {author} {\bibfnamefont{E.~R.}\ \bibnamefont{{Arriola}}},\ and\ \bibinfo
  {author} {\bibfnamefont{L.~L.}\ \bibnamefont{{Salcedo}}},\ }%
  \bibfield{journal}{%
  \Doi{10.1103/PhysRevD.74.065005}{\bibinfo {journal} {\prd}}\ }%
  \textbf{\bibinfo {volume} {74}},\ \bibinfo {eid} {065005} (\bibinfo {year}
  {2006}),\
  \Eprint{http://arxiv.org/abs/arXiv:hep-ph/0412308}{arXiv:hep-ph/0412308}%
  \bibAnnoteFile{NoStop}{Megias_06_A}%
\bibitem{Klevansky_92}%
  \BibitemOpen
  \bibfield{author}{%
  \bibinfo {author} {\bibfnamefont{S.}~\bibnamefont{Klevansky}},\ }%
  \bibfield{journal}{%
  \Doi{10.1103/RevModPhys.64.649}{\bibinfo {journal} {Rev.~Mod.~Phys.}}\ }%
  \textbf{\bibinfo {volume} {64}},\ \bibinfo {pages} {649} (\bibinfo {year}
  {1992})%
  \bibAnnoteFile{NoStop}{Klevansky_92}%
\bibitem{Buballa_05}%
  \BibitemOpen
  \bibfield{author}{%
  \bibinfo {author} {\bibfnamefont{M.}~\bibnamefont{{Buballa}}},\ }%
  \bibfield{journal}{%
  \Doi{10.1016/j.physrep.2004.11.004}{\bibinfo {journal} {Phys.~Rept.}}\ }%
  \textbf{\bibinfo {volume} {407}},\ \bibinfo {pages} {205} (\bibinfo {year}
  {2005}),\
  \Eprint{http://arxiv.org/abs/arXiv:hep-ph/0402234}{arXiv:hep-ph/0402234}%
  \bibAnnoteFile{NoStop}{Buballa_05}%
\bibitem{Costa_08}%
  \BibitemOpen
  \bibfield{author}{%
  \bibinfo {author} {\bibfnamefont{P.}~\bibnamefont{{Costa}}}, \bibinfo
  {author} {\bibfnamefont{M.~C.}\ \bibnamefont{{Ruivo}}},\ and\ \bibinfo
  {author} {\bibfnamefont{C.~A.}\ \bibnamefont{{de Sousa}}},\ }%
  \bibfield{journal}{%
  \Doi{10.1103/PhysRevD.77.096001}{\bibinfo {journal} {\prd}}\ }%
  \textbf{\bibinfo {volume} {77}},\ \bibinfo {eid} {096001} (\bibinfo {year}
  {2008}),\ \Eprint{http://arxiv.org/abs/0801.3417}{arXiv:0801.3417 [hep-ph]}%
  \bibAnnoteFile{NoStop}{Costa_08}%
\bibitem{Metzger_93}%
  \BibitemOpen
  \bibfield{author}{%
  \bibinfo {author} {\bibfnamefont{D.}~\bibnamefont{Metzger}}, \bibinfo
  {author} {\bibfnamefont{H.}~\bibnamefont{Meyer-Ortmanns}},\ and\ \bibinfo
  {author} {\bibfnamefont{H.}~\bibnamefont{Pirner}},\ }%
  \bibfield{journal}{%
  \Doi{10.1016/0370-2693(94)90328-X, 10.1016/0370-2693(94)90328-X}{\bibinfo
  {journal} {Phys.~Lett.}}\ }%
  \textbf{\bibinfo {volume} {B 321}},\ \bibinfo {pages} {66} (\bibinfo {year}
  {1994}),\
  \Eprint{http://arxiv.org/abs/arXiv:hep-ph/9312252}{arXiv:hep-ph/9312252}%
  \bibAnnoteFile{NoStop}{Metzger_93}%
\bibitem{Lenaghan_00}%
  \BibitemOpen
  \bibfield{author}{%
  \bibinfo {author} {\bibfnamefont{J.~T.}\ \bibnamefont{Lenaghan}}, \bibinfo
  {author} {\bibfnamefont{D.~H.}\ \bibnamefont{Rischke}},\ and\ \bibinfo
  {author} {\bibfnamefont{J.}~\bibnamefont{Schaffner-Bielich}},\ }%
  \bibfield{journal}{%
  \Doi{10.1103/PhysRevD.62.085008}{\bibinfo {journal} {prd}}\ }%
  \textbf{\bibinfo {volume} {62}},\ \bibinfo {pages} {085008} (\bibinfo {year}
  {2000}),\
  \Eprint{http://arxiv.org/abs/arXiv:nucl-th/0004006}{arXiv:nucl-th/0004006}%
  \bibAnnoteFile{NoStop}{Lenaghan_00}%
\bibitem{Scavenius_00}%
  \BibitemOpen
  \bibfield{author}{%
  \bibinfo {author} {\bibfnamefont{O.}~\bibnamefont{Scavenius}}, \bibinfo
  {author} {\bibfnamefont{A.}~\bibnamefont{Mocsy}}, \bibinfo {author}
  {\bibfnamefont{I.}~\bibnamefont{Mishustin}},\ and\ \bibinfo {author}
  {\bibfnamefont{D.}~\bibnamefont{Rischke}},\ }%
  \bibfield{journal}{%
  \Doi{10.1103/PhysRevC.64.045202}{\bibinfo {journal} {\prc}}\ }%
  \textbf{\bibinfo {volume} {64}},\ \bibinfo {pages} {045202} (\bibinfo {year}
  {2001}),\
  \Eprint{http://arxiv.org/abs/arXiv:nucl-th/0007030}{arXiv:nucl-th/0007030}%
  \bibAnnoteFile{NoStop}{Scavenius_00}%
\bibitem{Bowman_08}%
  \BibitemOpen
  \bibfield{author}{%
  \bibinfo {author} {\bibfnamefont{E.}~\bibnamefont{Bowman}}\ and\ \bibinfo
  {author} {\bibfnamefont{J.~I.}\ \bibnamefont{Kapusta}},\ }%
  \bibfield{journal}{%
  \Doi{10.1103/PhysRevC.79.015202}{\bibinfo {journal} {\prc}}\ }%
  \textbf{\bibinfo {volume} {79}},\ \bibinfo {pages} {015202} (\bibinfo {year}
  {2009}),\ \Eprint{http://arxiv.org/abs/0810.0042}{arXiv:0810.0042 [nucl-th]}%
  \bibAnnoteFile{NoStop}{Bowman_08}%
\bibitem{Schaefer_08}%
  \BibitemOpen
  \bibfield{author}{%
  \bibinfo {author} {\bibfnamefont{B.-J.}\ \bibnamefont{{Schaefer}}}\ and\
  \bibinfo {author} {\bibfnamefont{M.}~\bibnamefont{{Wagner}}},\ }%
  \bibfield{journal}{%
  \Doi{10.1103/PhysRevD.79.014018}{\bibinfo {journal} {\prd}}\ }%
  \textbf{\bibinfo {volume} {79}},\ \bibinfo {eid} {014018} (\bibinfo {year}
  {2009}),\ \Eprint{http://arxiv.org/abs/0808.1491}{arXiv:0808.1491 [hep-ph]}%
  \bibAnnoteFile{NoStop}{Schaefer_08}%
\bibitem{Polyakov_78}%
  \BibitemOpen
  \bibfield{author}{%
  \bibinfo {author} {\bibfnamefont{A.~M.}\ \bibnamefont{{Polyakov}}},\ }%
  \bibfield{journal}{%
  \Doi{10.1016/0370-2693(78)90737-2}{\bibinfo {journal} {Phys. Lett. B}}\ }%
  \textbf{\bibinfo {volume} {72}},\ \bibinfo {pages} {477} (\bibinfo {year}
  {1978})%
  \bibAnnoteFile{NoStop}{Polyakov_78}%
\bibitem{Marhauser_08}%
  \BibitemOpen
  \bibfield{author}{%
  \bibinfo {author} {\bibfnamefont{F.}~\bibnamefont{{Marhauser}}}\ and\
  \bibinfo {author} {\bibfnamefont{J.~M.}\ \bibnamefont{{Pawlowski}}},\ }%
  \bibfield{journal}{%
  \bibinfo {journal} {ArXiv e-prints}}%
   (\bibinfo {year} {2008}),\
  \Eprint{http://arxiv.org/abs/0812.1144}{arXiv:0812.1144 [hep-ph]}%
  \bibAnnoteFile{NoStop}{Marhauser_08}%
\bibitem{Megias_06_B}%
  \BibitemOpen
  \bibfield{author}{%
  \bibinfo {author} {\bibfnamefont{E.}~\bibnamefont{{Meg{\'{\i}}as}}}, \bibinfo
  {author} {\bibfnamefont{E.}~\bibnamefont{{Ruiz Arriola}}},\ and\ \bibinfo
  {author} {\bibfnamefont{L.~L.}\ \bibnamefont{{Salcedo}}},\ }%
  \bibfield{journal}{%
  \Doi{10.1088/1126-6708/2006/01/073}{\bibinfo {journal} {Journal of High
  Energy Physics}}\ }%
  \textbf{\bibinfo {volume} {1}},\ \bibinfo {eid} {073} (\bibinfo {year}
  {2006}),\
  \Eprint{http://arxiv.org/abs/arXiv:hep-ph/0505215}{arXiv:hep-ph/0505215}%
  \bibAnnoteFile{NoStop}{Megias_06_B}%
\bibitem{Braun:2007bx}%
  \BibitemOpen
  \bibfield{author}{%
  \bibinfo {author} {\bibfnamefont{J.}~\bibnamefont{{Braun}}}, \bibinfo
  {author} {\bibfnamefont{H.}~\bibnamefont{{Gies}}},\ and\ \bibinfo {author}
  {\bibfnamefont{J.~M.}\ \bibnamefont{{Pawlowski}}},\ }%
  \bibfield{journal}{%
  \Doi{10.1016/j.physletb.2010.01.009}{\bibinfo {journal} {Physics Letters B}}\
  }%
  \textbf{\bibinfo {volume} {684}},\ \bibinfo {pages} {262} (\bibinfo {year}
  {2010}),\ \Eprint{http://arxiv.org/abs/0708.2413}{arXiv:0708.2413 [hep-th]}%
  \bibAnnoteFile{NoStop}{Braun:2007bx}%
\bibitem{PDG_12}%
  \BibitemOpen
  \bibfield{author}{%
  \bibinfo {author} {\bibnamefont{{J. Beringer and Particle Data Group}}},\ }%
  \bibfield{journal}{%
  \Doi{10.1103/PhysRevD.86.010001}{\bibinfo {journal} {\prd}}\ }%
  \textbf{\bibinfo {volume} {86}},\ \bibinfo {eid} {010001} (\bibinfo {year}
  {2012})%
  \bibAnnoteFile{NoStop}{PDG_12}%
\bibitem{Boyd_96}%
  \BibitemOpen
  \bibfield{author}{%
  \bibinfo {author} {\bibfnamefont{G.}~\bibnamefont{{Boyd}}}, \bibinfo {author}
  {\bibfnamefont{J.}~\bibnamefont{{Engels}}}, \bibinfo {author}
  {\bibfnamefont{F.}~\bibnamefont{{Karsch}}}, \bibinfo {author}
  {\bibfnamefont{E.}~\bibnamefont{{Laermann}}}, \bibinfo {author}
  {\bibfnamefont{C.}~\bibnamefont{{Legeland}}}, \bibinfo {author}
  {\bibfnamefont{M.}~\bibnamefont{{L{\"u}tgemeier}}},\ and\ \bibinfo {author}
  {\bibfnamefont{B.}~\bibnamefont{{Petersson}}},\ }%
  \bibfield{journal}{%
  \Doi{10.1016/0550-3213(96)00170-8}{\bibinfo {journal} {Nucl. Phys. B}}\ }%
  \textbf{\bibinfo {volume} {469}},\ \bibinfo {pages} {419} (\bibinfo {year}
  {1996}),\
  \Eprint{http://arxiv.org/abs/arXiv:hep-lat/9602007}{arXiv:hep-lat/9602007}%
  \bibAnnoteFile{NoStop}{Boyd_96}%
\bibitem{Braun_06}%
  \BibitemOpen
  \bibfield{author}{%
  \bibinfo {author} {\bibfnamefont{J.}~\bibnamefont{{Braun}}}\ and\ \bibinfo
  {author} {\bibfnamefont{H.}~\bibnamefont{{Gies}}},\ }%
  \bibfield{journal}{%
  \Doi{10.1088/1126-6708/2006/06/024}{\bibinfo {journal} {JHEP}}\ }%
  \textbf{\bibinfo {volume} {6}},\ \bibinfo {pages} {24} (\bibinfo {year}
  {2006}),\
  \Eprint{http://arxiv.org/abs/arXiv:hep-ph/0602226}{arXiv:hep-ph/0602226}%
  \bibAnnoteFile{NoStop}{Braun_06}%
\bibitem{Braun_07}%
  \BibitemOpen
  \bibfield{author}{%
  \bibinfo {author} {\bibfnamefont{J.}~\bibnamefont{{Braun}}}\ and\ \bibinfo
  {author} {\bibfnamefont{H.}~\bibnamefont{{Gies}}},\ }%
  \bibfield{journal}{%
  \Doi{10.1016/j.physletb.2006.11.059}{\bibinfo {journal} {Phys. Lett. B}}\ }%
  \textbf{\bibinfo {volume} {645}},\ \bibinfo {pages} {53} (\bibinfo {year}
  {2007}),\
  \Eprint{http://arxiv.org/abs/arXiv:hep-ph/0512085}{arXiv:hep-ph/0512085}%
  \bibAnnoteFile{NoStop}{Braun_07}%
\bibitem{Braun_11}%
  \BibitemOpen
  \bibfield{author}{%
  \bibinfo {author} {\bibfnamefont{J.}~\bibnamefont{{Braun}}}, \bibinfo
  {author} {\bibfnamefont{L.~M.}\ \bibnamefont{{Haas}}}, \bibinfo {author}
  {\bibfnamefont{F.}~\bibnamefont{{Marhauser}}},\ and\ \bibinfo {author}
  {\bibfnamefont{J.~M.}\ \bibnamefont{{Pawlowski}}},\ }%
  \bibfield{journal}{%
  \Doi{10.1103/PhysRevLett.106.022002}{\bibinfo {journal} {Phys. Rev. Lett.}}\
  }%
  \textbf{\bibinfo {volume} {106}},\ \bibinfo {eid} {022002} (\bibinfo {year}
  {2011}),\ \Eprint{http://arxiv.org/abs/0908.0008}{arXiv:0908.0008 [hep-ph]}%
  \bibAnnoteFile{NoStop}{Braun_11}%
\bibitem{Pawlowski:2010ht}%
  \BibitemOpen
  \bibfield{author}{%
  \bibinfo {author} {\bibfnamefont{J.~M.}\ \bibnamefont{Pawlowski}},\ }%
  \bibfield{journal}{%
  \Doi{10.1063/1.3574945}{\bibinfo {journal} {AIP Conf.~Proc.}}\ }%
  \textbf{\bibinfo {volume} {1343}},\ \bibinfo {pages} {75} (\bibinfo {year}
  {2011}),\ \Eprint{http://arxiv.org/abs/1012.5075}{arXiv:1012.5075 [hep-ph]}%
  \bibAnnoteFile{NoStop}{Pawlowski:2010ht}%
\bibitem{Skokov_10}%
  \BibitemOpen
  \bibfield{author}{%
  \bibinfo {author} {\bibfnamefont{V.}~\bibnamefont{{Skokov}}}, \bibinfo
  {author} {\bibfnamefont{B.}~\bibnamefont{{Friman}}}, \bibinfo {author}
  {\bibfnamefont{E.}~\bibnamefont{{Nakano}}}, \bibinfo {author}
  {\bibfnamefont{K.}~\bibnamefont{{Redlich}}},\ and\ \bibinfo {author}
  {\bibfnamefont{B.-J.}\ \bibnamefont{{Schaefer}}},\ }%
  \bibfield{journal}{%
  \Doi{10.1103/PhysRevD.82.034029}{\bibinfo {journal} {\prd}}\ }%
  \textbf{\bibinfo {volume} {82}},\ \bibinfo {eid} {034029} (\bibinfo {year}
  {2010}),\ \Eprint{http://arxiv.org/abs/1005.3166}{arXiv:1005.3166 [hep-ph]}%
  \bibAnnoteFile{NoStop}{Skokov_10}%
\bibitem{Andersen_11}%
  \BibitemOpen
  \bibfield{author}{%
  \bibinfo {author} {\bibfnamefont{J.~O.}\ \bibnamefont{{Andersen}}}, \bibinfo
  {author} {\bibfnamefont{R.}~\bibnamefont{{Khan}}},\ and\ \bibinfo {author}
  {\bibfnamefont{L.~T.}\ \bibnamefont{{Kyllingstad}}},\ }%
  \bibfield{journal}{%
  \bibinfo {journal} {ArXiv e-prints}}%
   (\bibinfo {year} {2011}),\
  \Eprint{http://arxiv.org/abs/1102.2779}{arXiv:1102.2779 [hep-ph]}%
  \bibAnnoteFile{NoStop}{Andersen_11}%
\bibitem{MintzRamos_13}%
  \BibitemOpen
  \bibfield{author}{%
  \bibinfo {author} {\bibfnamefont{B.~W.}\ \bibnamefont{{Mintz}}}\ and\
  \bibinfo {author} {\bibfnamefont{R.~O.}\ \bibnamefont{{Ramos}}},\ }%
  \bibinfo {note} {work in progress}%
  \bibAnnoteFile{NoStop}{MintzRamos_13}%
\bibitem{Karsch_85}%
  \BibitemOpen
  \bibfield{author}{%
  \bibinfo {author} {\bibfnamefont{F.}~\bibnamefont{{Karsch}}}\ and\ \bibinfo
  {author} {\bibfnamefont{H.~W.}\ \bibnamefont{{Wyld}}},\ }%
  \bibfield{journal}{%
  \Doi{10.1103/PhysRevLett.55.2242}{\bibinfo {journal} {Phys. Rev. Lett.}}\ }%
  \textbf{\bibinfo {volume} {55}},\ \bibinfo {pages} {2242} (\bibinfo {year}
  {1985})%
  \bibAnnoteFile{NoStop}{Karsch_85}%
\bibitem{Langer:1969bc}%
  \BibitemOpen
  \bibfield{author}{%
  \bibinfo {author} {\bibfnamefont{J.~S.}\ \bibnamefont{Langer}},\ }%
  \bibfield{journal}{%
  \Doi{10.1016/0003-4916(69)90153-5}{\bibinfo {journal} {Ann. Phys.}}\ }%
  \textbf{\bibinfo {volume} {54}},\ \bibinfo {pages} {258} (\bibinfo {year}
  {1969})%
  \bibAnnoteFile{NoStop}{Langer:1969bc}%
\bibitem{Gunton_83}%
  \BibitemOpen
  \bibfield{author}{%
  \bibinfo {author} {\bibfnamefont{J.~D.}\ \bibnamefont{Gunton}}, \bibinfo
  {author} {\bibfnamefont{M.}~\bibnamefont{San~Miguel}},\ and\ \bibinfo
  {author} {\bibfnamefont{P.~S.}\ \bibnamefont{Sahni}},\ }%
  \bibfield{journal}{%
  \bibinfo {journal} {Phase Transitions and Critical Phenomena}\ }%
  \textbf{\bibinfo {volume} {8}},\ \bibinfo {pages} {267} (\bibinfo {year}
  {1983})%
  \bibAnnoteFile{NoStop}{Gunton_83}%
\bibitem{Cahn_58}%
  \BibitemOpen
  \bibfield{author}{%
  \bibinfo {author} {\bibfnamefont{J.~W.}\ \bibnamefont{{Cahn}}}\ and\ \bibinfo
  {author} {\bibfnamefont{J.~E.}\ \bibnamefont{{Hilliard}}},\ }%
  \bibfield{journal}{%
  \Doi{10.1063/1.1744102}{\bibinfo {journal} {\jcp}}\ }%
  \textbf{\bibinfo {volume} {28}},\ \bibinfo {pages} {258} (\bibinfo {year}
  {1958})%
  \bibAnnoteFile{NoStop}{Cahn_58}%
\bibitem{Cahn_59}%
  \BibitemOpen
  \bibfield{author}{%
  \bibinfo {author} {\bibfnamefont{J.~W.}\ \bibnamefont{{Cahn}}}\ and\ \bibinfo
  {author} {\bibfnamefont{J.~E.}\ \bibnamefont{{Hilliard}}},\ }%
  \bibfield{journal}{%
  \Doi{10.1063/1.1730447}{\bibinfo {journal} {\jcp}}\ }%
  \textbf{\bibinfo {volume} {31}},\ \bibinfo {pages} {688} (\bibinfo {year}
  {1959})%
  \bibAnnoteFile{NoStop}{Cahn_59}%
\bibitem{Coleman_78}%
  \BibitemOpen
  \bibfield{author}{%
  \bibinfo {author} {\bibfnamefont{S.}~\bibnamefont{{Coleman}}}, \bibinfo
  {author} {\bibfnamefont{V.}~\bibnamefont{{Glaser}}},\ and\ \bibinfo {author}
  {\bibfnamefont{A.}~\bibnamefont{{Martin}}},\ }%
  \bibfield{journal}{%
  \Doi{10.1007/BF01609421}{\bibinfo {journal} {Comm. Math. Phys.}}\ }%
  \textbf{\bibinfo {volume} {58}},\ \bibinfo {pages} {211} (\bibinfo {year}
  {1978})%
  \bibAnnoteFile{NoStop}{Coleman_78}%
\bibitem{Gleiser:1992ed}%
  \BibitemOpen
  \bibfield{author}{%
  \bibinfo {author} {\bibfnamefont{M.}~\bibnamefont{Gleiser}}\ and\ \bibinfo
  {author} {\bibfnamefont{R.~O.}\ \bibnamefont{Ramos}},\ }%
  \bibfield{journal}{%
  \Doi{10.1016/0370-2693(93)90365-O}{\bibinfo {journal} {Phys. Lett.}}\ }%
  \textbf{\bibinfo {volume} {B300}},\ \bibinfo {pages} {271} (\bibinfo {year}
  {1993}),\ \Eprint{http://arxiv.org/abs/hep-ph/9211219}{arXiv:hep-ph/9211219}%
  \bibAnnoteFile{NoStop}{Gleiser:1992ed}%
\bibitem{Ramos:1996at}%
  \BibitemOpen
  \bibfield{author}{%
  \bibinfo {author} {\bibfnamefont{R.~O.}\ \bibnamefont{Ramos}},\ }%
  \bibfield{journal}{%
  \Doi{10.1103/PhysRevD.54.4770}{\bibinfo {journal} {Phys. Rev.}}\ }%
  \textbf{\bibinfo {volume} {D54}},\ \bibinfo {pages} {4770} (\bibinfo {year}
  {1996}),\ \Eprint{http://arxiv.org/abs/hep-ph/9607417}{arXiv:hep-ph/9607417
  [hep-ph]}%
  \bibAnnoteFile{NoStop}{Ramos:1996at}%
\bibitem{Coleman_77}%
  \BibitemOpen
  \bibfield{author}{%
  \bibinfo {author} {\bibfnamefont{S.}~\bibnamefont{{Coleman}}},\ }%
  \bibfield{journal}{%
  \Doi{10.1103/PhysRevD.15.2929}{\bibinfo {journal} {\prd}}\ }%
  \textbf{\bibinfo {volume} {15}},\ \bibinfo {pages} {2929} (\bibinfo {year}
  {1977})%
  \bibAnnoteFile{NoStop}{Coleman_77}%
\bibitem{Coleman_77_erratum}%
  \BibitemOpen
  \bibfield{author}{%
  \bibinfo {author} {\bibfnamefont{S.}~\bibnamefont{{Coleman}}},\ }%
  \bibfield{journal}{%
  \Doi{10.1103/PhysRevD.16.1248}{\bibinfo {journal} {\prd}}\ }%
  \textbf{\bibinfo {volume} {16}},\ \bibinfo {pages} {1248} (\bibinfo {year}
  {1977})%
  \bibAnnoteFile{NoStop}{Coleman_77_erratum}%
\bibitem{Callan_77}%
  \BibitemOpen
  \bibfield{author}{%
  \bibinfo {author} {\bibfnamefont{C.~G.}\ \bibnamefont{{Callan}},
  \bibfnamefont{Jr.}}\ and\ \bibinfo {author}
  {\bibfnamefont{S.}~\bibnamefont{{Coleman}}},\ }%
  \bibfield{journal}{%
  \Doi{10.1103/PhysRevD.16.1762}{\bibinfo {journal} {\prd}}\ }%
  \textbf{\bibinfo {volume} {16}},\ \bibinfo {pages} {1762} (\bibinfo {year}
  {1977})%
  \bibAnnoteFile{NoStop}{Callan_77}%
\bibitem{Coleman_Book_88}%
  \BibitemOpen
  \bibfield{author}{%
  \bibinfo {author} {\bibfnamefont{S.}~\bibnamefont{{Coleman}}},\ }%
  \emph{\bibinfo {title} {{Aspects of Symmetry}}}\ (\bibinfo {publisher}
  {Cambridge University Press},\ \bibinfo {address} {Cambridge, England},\
  \bibinfo {year} {1988})%
  \bibAnnoteFile{NoStop}{Coleman_Book_88}%
\bibitem{Kashchiev_Book_00}%
  \BibitemOpen
  \bibfield{author}{%
  \bibinfo {author} {\bibfnamefont{D.}~\bibnamefont{Kashchiev}},\ }%
  \emph{\bibinfo {title} {Nucleation: Basic Theory with Applications}}\
  (\bibinfo {publisher} {Butterworth Heinemann, New York},\ \bibinfo {year}
  {2000})%
  \bibAnnoteFile{NoStop}{Kashchiev_Book_00}%
\bibitem{Vehkamaeki_Book_06}%
  \BibitemOpen
  \bibfield{author}{%
  \bibinfo {author} {\bibfnamefont{H.}~\bibnamefont{Vehkamaeki}},\ }%
  \emph{\bibinfo {title} {Classical Nucleation Theory in Multicomponent
  Systems}}\ (\bibinfo {publisher} {Springer, New York},\ \bibinfo {year}
  {2006})%
  \bibAnnoteFile{NoStop}{Vehkamaeki_Book_06}%
\bibitem{Linde_83}%
  \BibitemOpen
  \bibfield{author}{%
  \bibinfo {author} {\bibfnamefont{A.~D.}\ \bibnamefont{{Linde}}},\ }%
  \bibfield{journal}{%
  \Doi{10.1016/0550-3213(83)90293-6}{\bibinfo {journal} {Nucl. Phys. B}}\ }%
  \textbf{\bibinfo {volume} {216}},\ \bibinfo {pages} {421} (\bibinfo {year}
  {1983})%
  \bibAnnoteFile{NoStop}{Linde_83}%
\bibitem{Weinberg_93}%
  \BibitemOpen
  \bibfield{author}{%
  \bibinfo {author} {\bibfnamefont{E.~J.}\ \bibnamefont{{Weinberg}}},\ }%
  \bibfield{journal}{%
  \Doi{10.1103/PhysRevD.47.4614}{\bibinfo {journal} {\prd}}\ }%
  \textbf{\bibinfo {volume} {47}},\ \bibinfo {pages} {4614} (\bibinfo {year}
  {1993}),\
  \Eprint{http://arxiv.org/abs/arXiv:hep-ph/9211314}{arXiv:hep-ph/9211314}%
  \bibAnnoteFile{NoStop}{Weinberg_93}%
\bibitem{Pisarski_00}%
  \BibitemOpen
  \bibfield{author}{%
  \bibinfo {author} {\bibfnamefont{R.~D.}\ \bibnamefont{{Pisarski}}},\ }%
  \bibfield{journal}{%
  \Doi{10.1103/PhysRevD.62.111501}{\bibinfo {journal} {\prd}}\ }%
  \textbf{\bibinfo {volume} {62}},\ \bibinfo {eid} {111501} (\bibinfo {year}
  {2000}),\
  \Eprint{http://arxiv.org/abs/arXiv:hep-ph/0006205}{arXiv:hep-ph/0006205}%
  \bibAnnoteFile{NoStop}{Pisarski_00}%
\bibitem{Beinlich_97}%
  \BibitemOpen
  \bibfield{author}{%
  \bibinfo {author} {\bibfnamefont{B.}~\bibnamefont{{Beinlich}}}, \bibinfo
  {author} {\bibfnamefont{F.}~\bibnamefont{{Karsch}}},\ and\ \bibinfo {author}
  {\bibfnamefont{A.}~\bibnamefont{{Peikert}}},\ }%
  \bibfield{journal}{%
  \Doi{10.1016/S0370-2693(96)01401-3}{\bibinfo {journal} {Phys.~Lett.B}}\ }%
  \textbf{\bibinfo {volume} {390}},\ \bibinfo {pages} {268} (\bibinfo {year}
  {1997}),\
  \Eprint{http://arxiv.org/abs/arXiv:hep-lat/9608141}{arXiv:hep-lat/9608141}%
  \bibAnnoteFile{NoStop}{Beinlich_97}%
\bibitem{Lucini_05}%
  \BibitemOpen
  \bibfield{author}{%
  \bibinfo {author} {\bibfnamefont{B.}~\bibnamefont{{Lucini}}}, \bibinfo
  {author} {\bibfnamefont{M.}~\bibnamefont{{Teper}}},\ and\ \bibinfo {author}
  {\bibfnamefont{U.}~\bibnamefont{{Wenger}}},\ }%
  \bibfield{journal}{%
  \Doi{10.1088/1126-6708/2005/02/033}{\bibinfo {journal} {Journal of High
  Energy Physics}}\ }%
  \textbf{\bibinfo {volume} {2}},\ \bibinfo {eid} {033} (\bibinfo {month}
  {Feb.}\ \bibinfo {year} {2005}),\
  \Eprint{http://arxiv.org/abs/arXiv:hep-lat/0502003}{arXiv:hep-lat/0502003}%
  \bibAnnoteFile{NoStop}{Lucini_05}%
\bibitem{Dumitru_12}%
  \BibitemOpen
  \bibfield{author}{%
  \bibinfo {author} {\bibfnamefont{A.}~\bibnamefont{{Dumitru}}}, \bibinfo
  {author} {\bibfnamefont{Y.}~\bibnamefont{{Guo}}}, \bibinfo {author}
  {\bibfnamefont{Y.}~\bibnamefont{{Hidaka}}}, \bibinfo {author}
  {\bibfnamefont{C.~P.}\ \bibnamefont{{Korthals Altes}}},\ and\ \bibinfo
  {author} {\bibfnamefont{R.~D.}\ \bibnamefont{{Pisarski}}},\ }%
  \bibfield{journal}{%
  \bibinfo {journal} {ArXiv e-prints}}%
   (\bibinfo {month} {May}\ \bibinfo {year} {2012}),\
  \Eprint{http://arxiv.org/abs/1205.0137}{arXiv:1205.0137 [hep-ph]}%
  \bibAnnoteFile{NoStop}{Dumitru_12}%
\bibitem{Rajaraman_Book_87}%
  \BibitemOpen
  \bibfield{author}{%
  \bibinfo {author} {\bibfnamefont{R.}~\bibnamefont{Rajaraman}},\ }%
  \emph{\bibinfo {title} {{An Introduction to Solitons and Instantons in
  Quantum Field Theory}}}\ (\bibinfo {publisher} {North-Holland, Amsterdam},\
  \bibinfo {year} {1987})%
  \bibAnnoteFile{NoStop}{Rajaraman_Book_87}%
\bibitem{Kaczmarek_02}%
  \BibitemOpen
  \bibfield{author}{%
  \bibinfo {author} {\bibfnamefont{O.}~\bibnamefont{{Kaczmarek}}}, \bibinfo
  {author} {\bibfnamefont{F.}~\bibnamefont{{Karsch}}}, \bibinfo {author}
  {\bibfnamefont{P.}~\bibnamefont{{Petreczky}}},\ and\ \bibinfo {author}
  {\bibfnamefont{F.}~\bibnamefont{{Zantow}}},\ }%
  \bibfield{journal}{%
  \Doi{10.1016/S0370-2693(02)02415-2}{\bibinfo {journal} {Physics Letters B}}\
  }%
  \textbf{\bibinfo {volume} {543}},\ \bibinfo {pages} {41} (\bibinfo {year}
  {2002}),\
  \Eprint{http://arxiv.org/abs/arXiv:hep-lat/0207002}{arXiv:hep-lat/0207002}%
  \bibAnnoteFile{NoStop}{Kaczmarek_02}%
\bibitem{Mykkanen_12}%
  \BibitemOpen
  \bibfield{author}{%
  \bibinfo {author} {\bibfnamefont{A.}~\bibnamefont{{Mykk{\"a}nen}}}, \bibinfo
  {author} {\bibfnamefont{M.}~\bibnamefont{{Panero}}},\ and\ \bibinfo {author}
  {\bibfnamefont{K.}~\bibnamefont{{Rummukainen}}},\ }%
  \bibfield{journal}{%
  \Doi{10.1007/JHEP05(2012)069}{\bibinfo {journal} {Journal of High Energy
  Physics}}\ }%
  \textbf{\bibinfo {volume} {5}},\ \bibinfo {pages} {69} (\bibinfo {year}
  {2012}),\ \Eprint{http://arxiv.org/abs/1202.2762}{arXiv:1202.2762 [hep-lat]}%
  \bibAnnoteFile{NoStop}{Mykkanen_12}%
\bibitem{Weber_05}%
  \BibitemOpen
  \bibfield{author}{%
  \bibinfo {author} {\bibfnamefont{F.}~\bibnamefont{{Weber}}},\ }%
  \bibfield{journal}{%
  \Doi{10.1016/j.ppnp.2004.07.001}{\bibinfo {journal} {Progress in Particle and
  Nuclear Physics}}\ }%
  \textbf{\bibinfo {volume} {54}},\ \bibinfo {pages} {193} (\bibinfo {year}
  {2005}),\
  \Eprint{http://arxiv.org/abs/arXiv:astro-ph/0407155}{arXiv:astro-ph/0407155}%
  \bibAnnoteFile{NoStop}{Weber_05}%
\bibitem{Sagert_09}%
  \BibitemOpen
  \bibfield{author}{%
  \bibinfo {author} {\bibfnamefont{I.}~\bibnamefont{{Sagert}}}, \bibinfo
  {author} {\bibfnamefont{T.}~\bibnamefont{{Fischer}}}, \bibinfo {author}
  {\bibfnamefont{M.}~\bibnamefont{{Hempel}}}, \bibinfo {author}
  {\bibfnamefont{G.}~\bibnamefont{{Pagliara}}}, \bibinfo {author}
  {\bibfnamefont{J.}~\bibnamefont{{Schaffner-Bielich}}}, \bibinfo {author}
  {\bibfnamefont{A.}~\bibnamefont{{Mezzacappa}}}, \bibinfo {author}
  {\bibfnamefont{F.-K.}\ \bibnamefont{{Thielemann}}},\ and\ \bibinfo {author}
  {\bibfnamefont{M.}~\bibnamefont{{Liebend{\"o}rfer}}},\ }%
  \bibfield{journal}{%
  \Doi{10.1103/PhysRevLett.102.081101}{\bibinfo {journal} {Physical Review
  Letters}}\ }%
  \textbf{\bibinfo {volume} {102}},\ \bibinfo {eid} {081101} (\bibinfo {year}
  {2009}),\ \Eprint{http://arxiv.org/abs/0809.4225}{arXiv:0809.4225
  [astro-ph]}%
  \bibAnnoteFile{NoStop}{Sagert_09}%
\bibitem{Boyanovsky_06}%
  \BibitemOpen
  \bibfield{author}{%
  \bibinfo {author} {\bibfnamefont{D.}~\bibnamefont{{Boyanovsky}}}, \bibinfo
  {author} {\bibfnamefont{H.~J.}\ \bibnamefont{{de Vega}}},\ and\ \bibinfo
  {author} {\bibfnamefont{D.~J.}\ \bibnamefont{{Schwarz}}},\ }%
  \bibfield{journal}{%
  \Doi{10.1146/annurev.nucl.56.080805.140539}{\bibinfo {journal} {Annual Review
  of Nuclear and Particle Science}}\ }%
  \textbf{\bibinfo {volume} {56}},\ \bibinfo {pages} {441} (\bibinfo {year}
  {2006}),\
  \Eprint{http://arxiv.org/abs/arXiv:hep-ph/0602002}{arXiv:hep-ph/0602002}%
  \bibAnnoteFile{NoStop}{Boyanovsky_06}%
\bibitem{Schettler_11}%
  \BibitemOpen
  \bibfield{author}{%
  \bibinfo {author} {\bibfnamefont{S.}~\bibnamefont{{Schettler}}}, \bibinfo
  {author} {\bibfnamefont{T.}~\bibnamefont{{Boeckel}}},\ and\ \bibinfo {author}
  {\bibfnamefont{J.}~\bibnamefont{{Schaffner-Bielich}}},\ }%
  \bibfield{journal}{%
  \Doi{10.1103/PhysRevD.83.064030}{\bibinfo {journal} {\prd}}\ }%
  \textbf{\bibinfo {volume} {83}},\ \bibinfo {eid} {064030} (\bibinfo {year}
  {2011}),\ \Eprint{http://arxiv.org/abs/1010.4857}{arXiv:1010.4857
  [astro-ph.CO]}%
  \bibAnnoteFile{NoStop}{Schettler_11}%
\bibitem{Schwarz_09}%
  \BibitemOpen
  \bibfield{author}{%
  \bibinfo {author} {\bibfnamefont{D.~J.}\ \bibnamefont{{Schwarz}}}\ and\
  \bibinfo {author} {\bibfnamefont{M.}~\bibnamefont{{Stuke}}},\ }%
  \bibfield{journal}{%
  \Doi{10.1088/1475-7516/2009/11/025}{\bibinfo {journal} {\jcap}}\ }%
  \textbf{\bibinfo {volume} {11}},\ \bibinfo {eid} {025} (\bibinfo {year}
  {2009}),\ \Eprint{http://arxiv.org/abs/0906.3434}{arXiv:0906.3434 [hep-ph]}%
  \bibAnnoteFile{NoStop}{Schwarz_09}%
\bibitem{Randrup_09}%
  \BibitemOpen
  \bibfield{author}{%
  \bibinfo {author} {\bibfnamefont{J.}~\bibnamefont{{Randrup}}},\ }%
  \bibfield{journal}{%
  \Doi{10.1103/PhysRevC.79.054911}{\bibinfo {journal} {\prc}}\ }%
  \textbf{\bibinfo {volume} {79}},\ \bibinfo {eid} {054911} (\bibinfo {year}
  {2009}),\ \Eprint{http://arxiv.org/abs/0903.4736}{arXiv:0903.4736 [nucl-th]}%
  \bibAnnoteFile{NoStop}{Randrup_09}%
\bibitem{Steinheimer_12}%
  \BibitemOpen
  \bibfield{author}{%
  \bibinfo {author} {\bibfnamefont{J.}~\bibnamefont{{Steinheimer}}}\ and\
  \bibinfo {author} {\bibfnamefont{J.}~\bibnamefont{{Randrup}}},\ }%
  \bibfield{journal}{%
  \bibinfo {journal} {Phys.~Rev.~Lett.}\ }%
  \textbf{\bibinfo {volume} {109}},\ \bibinfo {pages} {212301} (\bibinfo {year}
  {2012}),\ \Eprint{http://arxiv.org/abs/1209.2462}{arXiv:1209.2462 [nucl-th]}%
  \bibAnnoteFile{NoStop}{Steinheimer_12}%
\bibitem{Kapusta_95}%
  \BibitemOpen
  \bibfield{author}{%
  \bibinfo {author} {\bibfnamefont{J.~I.}\ \bibnamefont{{Kapusta}}}, \bibinfo
  {author} {\bibfnamefont{A.~P.}\ \bibnamefont{{Vischer}}},\ and\ \bibinfo
  {author} {\bibfnamefont{R.}~\bibnamefont{{Venugopalan}}},\ }%
  \bibfield{journal}{%
  \Doi{10.1103/PhysRevC.51.901}{\bibinfo {journal} {\prc}}\ }%
  \textbf{\bibinfo {volume} {51}},\ \bibinfo {pages} {901} (\bibinfo {year}
  {1995}),\
  \Eprint{http://arxiv.org/abs/arXiv:nucl-th/9408029}{arXiv:nucl-th/9408029}%
  \bibAnnoteFile{NoStop}{Kapusta_95}%
\bibitem{Csernai_95}%
  \BibitemOpen
  \bibfield{author}{%
  \bibinfo {author} {\bibfnamefont{L.~P.}\ \bibnamefont{{Csernai}}}\ and\
  \bibinfo {author} {\bibfnamefont{I.~N.}\ \bibnamefont{{Mishustin}}},\ }%
  \bibfield{journal}{%
  \Doi{10.1103/PhysRevLett.74.5005}{\bibinfo {journal} {Phys.~Rev.~Lett.}}\ }%
  \textbf{\bibinfo {volume} {74}},\ \bibinfo {pages} {5005} (\bibinfo {year}
  {1995})%
  \bibAnnoteFile{NoStop}{Csernai_95}%
\end{thebibliography}%

\end{document}